\documentclass[acmsmall,screen]{acmart}

\AtBeginDocument{%
  }

\usepackage{color,soul}
\usepackage{subcaption}
\usepackage{algpseudocode}
\usepackage{algorithm}
\usepackage[framemethod=tikz]{mdframed}
\begin{document}

\title{Optimization is Better than Generation: Optimizing Commit Message Leveraging Human-written Commit Message}

\author{Jiawei Li}
\orcid{0000-0002-4434-4812}
\affiliation{%
  \institution{University of California, Irvine}
  \city{Irvine}
  \country{USA}
}
\email{jiawl28@uci.edu}

\author{David Faragó}
\orcid{0009-0006-2380-6076}
\affiliation{%
  \institution{Innoopract}
  \city{}
  \country{Germany}
}
\affiliation{%
  \institution{QPR Technologies}
  \city{}
  \country{Germany}
}
\email{farago@qpr-technologies.de}

\author{Christian Petrov}
\orcid{0000-0001-8776-4289}
\affiliation{%
  \institution{Innoopract}
  \city{}
  \country{Germany}
}
\email{cpetrov@innoopract.com}

\author{Iftekhar Ahmed}
\orcid{0000-0001-8221-5352}
\affiliation{%
  \institution{University of California, Irvine}
  \city{Irvine}
  \country{USA}
}
\email{iftekha@uci.edu}



\begin{abstract}
Commit messages are crucial in software development, supporting maintenance tasks and communication among developers. While Large Language Models (LLMs) have advanced Commit Message Generation (CMG) using various software contexts, some contexts developers consider are often missed by CMG techniques and can’t be easily retrieved or even retrieved at all by automated tools. To address this, we propose Commit Message Optimization (CMO), which enhances human-written messages by leveraging LLMs and search-based optimization. CMO starts with human-written messages and iteratively improves them by integrating key contexts and feedback from external evaluators. Our extensive evaluation shows CMO generates commit messages that are significantly more Rational, Comprehensive, and Expressive while outperforming state-of-the-art CMG methods and human messages 88.2\%-95.4\% of the time.
\end{abstract}





\maketitle

\section{Introduction}
\label{sec:intro}

% Commit message is important, high quality means what, why and context and more. However, developers don't write them due to various factors, leaving commit message quality at risk.

Commit messages are essential for explaining code changes, including what was changed and why, as well as providing context that aids software maintenance. High-quality messages concisely describe the changes (``What" information), the rationale behind them (``Why" information), and additional relevant information, which is critical for understanding the impact on the codebase \cite{tian2022makes,li2024only}. Moreover, well-written commit messages can reduce software defect proneness \cite{li2023commit}. However, developers often neglect this practice due to time or motivation constraints, resulting in vague or incomplete messages, even in projects like those under Apache Software Foundation \cite{tian2022makes, li2023commit}. Dyer et al. \cite{dyer2013boa} found that around 14\% of commit messages in over 23,000 SourceForge projects were left blank, further highlighting this issue. %Addressing this problem is vital for improving software quality and maintainability.

%OLD: Software developers write commit messages in natural language to explain the code changes that they have committed to the software repository, aiming to help code change understanding \cite{tao2012software}. High-quality commit messages concisely provide a summary of what has been changed (the ``What" information), the motivations or justifications behind the changes (the ``Why" information) \cite{tian2022makes}, the maintenance type of the code changes, and any additional context that would aid code comprehension \cite{li2024only}, which is essential for software maintenance. In addition, the quality of the commit messages has also been shown to have an impact on software defect proneness \cite{li2023commit}. Nonetheless, due to a lack of time, motivation, and effort, developers do not always write high-quality commit messages, even in well-established organizations such as Apache Software Foundation \cite{tian2022makes, li2023commit}. Also, according to Dyer et al. \cite{dyer2013boa}, about 14\% of the commit messages in more than 23,000 projects on SourceForge were left empty.

%\hl{remove some of the reference in the below para, just keep few.  Jiawei: done}

% traditional CMG is designed to help developers write messages, but the generated messages have significant quality issues.

To assist developers in writing commit messages and elevate the commit message quality issue, numerous automated Commit Message Generation (CMG) techniques have been proposed by the Software Engineering (SE) research community. These methods, given a set of code changes, aim to generate descriptive commit messages. These techniques can be broadly categorized into \textit{Template-based} approaches where predefined patterns are used for generating messages \cite{linares2015changescribe, shen2016automatic}.
\textit{Retrieval-based} methods which find similar past commit messages and adapt them \cite{huang2020learning, shi2022race}.
\textit{Translation-based} techniques treat commit message generation as a language translation problem, translating code changes into natural language \cite{xu2019commit,liu2020atom, nie2021coregen}.
\textit{Hybrid} approaches combine elements from the above techniques to enhance performance \cite{liu2020atom, shi2022race,he2023come}. However, researchers have noted that these methods often produce messages lacking key information, such as the ``What'' or ``Why,'' provide insufficient context, follow simple patterns, or are semantically unrelated to the human-written messages and difficult to read \cite{jiang2017automatically,wang2021quality,liu2018neural,dong2023revisiting}. This underscores the need for further improvements in CMG techniques to generate higher-quality messages.

%OLD: To assist developers in writing commit messages, numerous automated Commit Message Generation (CMG) techniques have been proposed by the Software Engineering (SE) research community. Given the code changes in a commit, CMG aims to automatically generate a descriptive commit message. These techniques can be broadly categorized into template-based \cite{buse2010automatically, cortes2014automatically, linares2015changescribe, shen2016automatic}, retrieval-based \cite{huang2017mining, liu2018neural, huang2020learning, shi2022race}, translation-based \cite{jiang2017automatically, loyola2017neural, xu2019commit, liu2019generating, liu2020atom, nie2021coregen}, and hybrid approaches \cite{liu2020atom, wang2021context, shi2022race,he2023come}. However, researchers have found numerous generated commit messages that miss essential information (i.e., ``What" or ``Why"), lack sufficient context information, follow simple patterns, are semantically unrelated to the human-written messages and poorly readable \cite{jiang2017automatically,wang2021quality,liu2018neural,dong2023revisiting}. 

% OMG is proposed to solve these quality problems these CMG techniques have. However, it also has limitations such as not considering enough nuances and context info. We posit we can consider more context that developers consider when writing the messages.

Recent advancements in Large Language Models (LLMs) have dramatically improved the performance of various software engineering tasks, including CMG. A notable state-of-the-art technique was proposed by Li et al. \cite{li2024only}, leveraging GPT-4 with the ReAct prompting technique \cite{yao2022react}. Their method named Omniscient Message Generator (OMG), achieved superior results by incorporating more contextual information than other CMG approaches. While most CMG techniques only use the code changes without additional context \cite{liu2020atom, shi2022race, dong2022fira}, OMG integrates various software contexts, such as pull request titles and method/class summaries. Since commit messages serve as one of the primary communication channels for developers facilitating discussions on diverse software maintenance activities that often depend on multiple types of context \cite{mannan2020relationship,tian2022makes,li2023commit}, it is plausible that developers use even more nuanced contextual information than what OMG considers. This raises the question:

\textbf{RQ1: What software context information do developers consider when writing commit messages that the CMG techniques have missed?}

% However, not all information can be readily retrieved with tools. Developers also consider different contexts for different diffs. Moreover, long prompt can hurt LLMs' performance. Thus, human's guidance is needed.

Developers often use a variety of information sources during software maintenance, such as code, issue reports, and interpersonal communications \cite{tian2022makes}. This variety makes it difficult for automated tools to capture all the necessary human-considered contexts for generating high-quality commit messages. To address this, integrating human guidance is essential for producing commit messages that reflect these diverse contexts. Human guidance has already proven effective in tasks like code repair and assertion generation \cite{bohme2020human,zamprogno2022dynamic}, and similar methods could be beneficial in Commit Message Generation (CMG). Eliseeva et al. \cite{eliseeva2023commit} introduced Commit Message Completion (CMC) by using human-typed prefixes to help LLMs generate more accurate commit messages. However, their study did not evaluate whether the generated messages included essential information like the ``What''/``Why'' information or included maintenance type/additional contexts, nor did they consider other important sources of context that affect message quality. Furthermore, they only considered the preceding commit to assist in message personalization, overlooking other critical software contexts that can impact commit message quality \cite{tian2022makes, li2023commit, li2024only}.

Inspired by Commit Message Completion (CMC), we propose a new method called \textit{Commit Message Optimization} (CMO) to enhance the quality of commit messages by leveraging a search-based optimization approach. Starting with a human-written message, CMO iteratively refines and improves the message to optimize its quality. This method addresses the fact that human-written commit messages can often contain quality issues, such as missing essential information or poor readability \cite{tian2022makes, li2023commit}. CMO brings two key advantages; firstly, by starting from a human-written message, the model gains access to the nuanced context a developer would naturally consider for the associated code changes. This includes information that might be impractical for automated systems to retrieve but is readily available in a developer's mind. Second, using the power of LLMs, CMO can ``fix'' or optimize the commit message by incorporating widely considered contexts and addressing quality issues. The iterative process allows the LLM to enhance the message with more relevant information, improving its clarity and completeness.

Given the diverse preferences developers show in describing software maintenance activities \cite{tian2022makes}, it becomes critical to recognize that the relevance of specific software contexts can vary based on the nature and motivations behind the code changes. Including all available contexts indiscriminately might dilute the quality of commit messages, as Large Language Models (LLMs) tend to perform worse when confronted with irrelevant information in the prompt \cite{jones2022capturing,shi2023large}. Moreover, approaches like OMG, which depend entirely on GPT-4 for context reasoning and message generation, face limitations because LLMs often make mistakes in reasoning and judgment \cite{stechly2023gpt,valmeekam2023planning,zhang2023language}. To counter this, instead of relying solely on GPT-4's reasoning, we prompt the model to focus on a single type of software context at each optimization step—contexts that developers frequently reference. Additionally, we introduced external evaluators to provide feedback on the commit message quality, using this feedback as an explicit objective function. This guided optimization improves the precision and relevance of the generated commit messages, overcoming the reliance on LLM's potentially flawed reasoning and ensuring the inclusion of only the most pertinent contexts.

To assess the effectiveness of our approach, we propose the following research questions:

\textbf{RQ2: How does our approach perform compared to the state-of-the-art CMG technique, CMC technique, and human developers?}

\textbf{RQ3: How do software context collection tools and search-based optimization contribute to the overall effectiveness?}

% to answer the above two rqs, we did ...
Since our approach has a number of adjustable hyper-parameters, we select the best best-performing hyper-parameters by manually checking the optimized messages, and use the approach with the best-performing hyper-parameters to compare with the baselines (RQ2) and the variants of our approach without certain components (RQ3). %\hl{Jiawei: this part may need some change based on how much we emphasize on the hyper-parameter tuning. I'll update later.} 

%Specifically, we perform all the experiments on the dataset adopted by OMG \cite{li2024only} that contains code changes and human-written commit messages from 32 Apache projects, along with OMG-generated messages. The results show that the commit messages optimized by CMO have significantly higher quality than the original human-written messages in terms of the evaluation metrics adopted in OMG. Compared with the messages generated by OMG and completed by CMC, our approach can consistently generate better messages according to humans. Similar to the findings in \cite{li2024only} that widely-used automatic metrics such as BLEU are questionable in evaluating CMG, these metrics do not capture the quality improvement in the optimized commit messages. 

% Finally, we also explore whether the components in CMO can be applied to existing CMG and CMC techniques to improve the quality of the generated/completed messages, which prompts us to ask our last research question:

% \textbf{RQ4: How can software context collection tools and search-based optimization help the state-of-the-art CMG technique and CMC technique?}

To sum up, our contributions are as follows:
\begin{itemize}
\item We systematically identify \textbf{software contexts that are widely considered by developers but missed by existing CMG approaches}. 
%\hl{Jiawei: ``Consider" is a strong word and we didn't survey developers, do we change this word?}
\item We propose a novel framework, \textbf{CMO, that optimizes existing human-written commit messages} using LLMs, software contexts, and automated evaluators.
\item Our human evaluation shows that our approach with appropriate hyper-parameters can \textbf{significantly improve human-written commit message quality and outperform existing CMG and CMC techniques}.
% \item The major components in CMO can \textbf{enhance the performance of the state-of-the-art CMG technique and CMC technique}. 
\end{itemize}

The remainder of this paper is organized as follows: In Section \ref{sec:rw}, we present the background and related works. We outline the methodology for identifying the missing contexts and the context themes in Section \ref{sec:rq1}. Next, we introduce the experimental setup for answering our RQ2 and RQ3 in Section \ref{sec:setup}. Then, the motivation and design of CMO are described in Section \ref{sec:cmo}. Section \ref{sec:rq2} and \ref{sec:rq3} provide the evaluation results. Section \ref{sec:ttv} shows potential threats to validity of our findings. Finally, we conclude the findings of our study in Section \ref{sec:conclusion}.

\section{Background \& Related Work}
\label{sec:rw}

\subsection{Commit Message Quality}
Researchers have traditionally assessed commit message quality using syntactic features like message length, word frequency, punctuation, and imperative verb mood~\cite{chahal2018developer, chen2020project}, but these methods overlook the semantic content. To address this, Tian et al.\cite{tian2022makes} proposed that high-quality commit messages should summarize code changes (``What'') and explain their motivations (``Why''). However, they considered issue report/pull request links as providing the ``Why'' without evaluating the content of these links. Li et al.~\cite{li2023commit} addressed this by considering both commit messages and their link contents, while in a separate study, Li et al.~\cite{li2024only} expanded the criteria to include additional quality expectations from developers. Our study aims to enhance existing human-written commit messages by optimizing them according to these quality factors.

\subsection{Commit Message Generation}
Composing a high-quality commit message that comprehensively explains code changes can be challenging for developers, particularly when time and motivation are limited. To tackle this, researchers have developed various CMG techniques that automate message generation from code changes (i.e., git diff). These techniques have a variety of underlying mechanisms. Template-based approaches~\cite{linares2015changescribe, shen2016automatic} use predefined templates for specific code changes but struggle with generalization and often omit the crucial ``Why'' information. Retrieval-based methods~\cite{huang2020learning, shi2022race} retrieve similar code changes and reuse their commit messages. However, this approach relies on human-written messages, which are frequently flawed—about 44\% of the commit messages in Open-Source Software (OSS) projects miss critical ``What'' or ``Why'' details~\cite{tian2022makes}. Translation-based techniques~\cite{liu2020atom, nie2021coregen} use Neural Machine Translation (NMT) to ``translate'' code changes into commit messages but a majority (90\%) of the generated messages are short, simple, and inadequate~\cite{dong2023revisiting}. Hybrid approaches~\cite{shi2022race, he2023come} combine retrieval and translation methods, yet their effectiveness can be hampered by the limitations of both techniques and issues with the fusion process~\cite{he2023come}.

Despite advancements in CMG, most approaches still rely solely on the git diff as input, overlooking associated software contexts that could enhance message quality~\cite{liu2020atom, shi2022race, dong2022fira}. Only a few studies have incorporated additional contexts, such as issue states or modified ASTs, alongside code changes~\cite{liu2020atom, tao2021evaluation, wang2023delving}. Recently, Eliseeva et al.\cite{eliseeva2023commit} introduced CMC that uses the prefix of human-written messages and commit history to improve performance and tries to address the issue of generating commit messages that are semantically irrelevant to human-written ones\cite{jiang2017automatically, liu2018neural,wang2021quality}. However, the authors did not assess the semantic quality of completed messages~\cite{tian2022makes,li2023commit,li2024only} or consider essential software contexts crucial for producing high-quality commit messages~\cite{li2024only}.

%OLD: Despite continuous technical advances in CMG, most approaches still only take git diff as the sole input while associated software contexts have the potential to improve the generated commit message quality \cite{jiang2017automatically, liu2018neural, loyola2017neural, xu2019commit}. Only a few studies have utilized contexts such as issue states or modified ASTs along with the code changes \cite{liu2020atom, tao2021evaluation, wang2023delving}. More recently, Eliseeva et al. \cite{eliseeva2023commit} proposed CMC with the goal of tackling the issue where existing CMG approaches generate messages that are semantically irrelevant to the reference human-written messages \cite{jiang2017automatically, liu2018neural,wang2021quality}. In CMC, the inputs include both the prefix of human-written messages and developers' commit message history, which helped the performance of applied CMG approaches. Nonetheless, the authors did not evaluate the semantic quality of the completed messages \cite{tian2022makes,li2023commit,li2024only} nor consider software contexts that are essential for generating high-quality commit messages \cite{li2024only}.

With the advent of powerful LLMs such as ChatGPT~\cite{chatgpt} and GPT-4~\cite{gpt-4}, researchers have integrated these models into CMG techniques, achieving results comparable or superior to existing approaches~\cite{li2024only,wu2024commit}. To meet software developers' expectations for high-quality commit messages while considering broader software contexts, Li et al.\cite{li2024only} introduced OMG, which uses ReAct prompting\cite{yao2022react} with GPT-4. This approach retrieves relevant software contexts, such as issue reports, pull requests, and method/class changes, to ground commit messages in factual software repository data. Human evaluations showed that OMG had set a new state-of-the-art CMG, with some messages deemed better than those that humans wrote. However, OMG still missed several software context details that humans commonly consider when writing commit messages. To address this, our study developed additional retrieval tools and adopted the human-written commit message as a ``prefix'', aiming to enhance its quality, ensuring critical information is not overlooked.

%OLD: With the advent of powerful LLMs such as ChatGPT \cite{chatgpt} and GPT-4 \cite{gpt-4}, researchers have started to integrate them into CMG techniques and achieved better or comparable performance than existing approaches \cite{tao2024kadel,li2024only,wu2024commit}. To generate commit messages that satisfy software developers' expectations for commit message quality and consider even more software contexts than previous works, Li et al. \cite{li2024only} proposed OMG where they employed ReAct prompting \cite{yao2022react} with GPT-4 to generate commit messages by allowing it to engage in active reasoning with retrieved software contexts. In this approach, the authors built tools to retrieve relevant software contexts including issue report/pull request and changed method/class, aiming to make the generated messages more factually grounded on the software repositories and more comprehensive with sufficient information. Human evaluation showed OMG achieves a new state-of-the-art performance in CMG and it even generates messages that are considered better than human-written ones. However, we found OMG still missed several types of software context information that have been widely considered by humans when writing commit messages. Thus, in this study, we built extra retrieval tools and adopt the whole human-written commit message as a ``prefix'' with the goal of augmenting it to be of the highest quality rather than generating a message from scratch where important human-considered information is overlooked.  

\subsection{Large Language Models \& Search-based Optimization}
In recent years, Large Language Models (LLMs) have been widely adopted to automate a range of tasks across various domains, including Software Engineering (SE) \cite{li2024only, wu2024commit}. These tasks often involve both source code and natural language, such as Commit Message Generation (CMG)\cite{li2024only, wu2024commit}, code summarization~\cite{geng2024large, ahmed2024automatic, virk2024enhancing}, and code generation~\cite{jiang2023self, li2023skcoder, mu2024clarifygpt}. LLMs are also increasingly used in optimization, search-based planning, and problem-solving across domains~\cite{liu2023large,yao2024tree, yang2024large}. An optimization typically starts from an initial solution, then iteratively updates the existing solutions or generates new solutions based on the existing ones to optimize the objective function. Instead of carefully designing the optimization problem and relying on an external programmed solver to produce update steps, including the natural language optimization problem in the prompt and instructing the LLMs to iteratively generate better solutions based on problem description and previously generated solutions have achieved decent performance \cite{yao2024tree,yang2024large}. However, none of the existing CMG techniques have explored the applicability of this approach. We aim to do so in this paper. We leverage a state-of-the-art LLM as an optimizer to enhance the quality of generated commit messages. In our approach, each optimization step refines the previously generated message, incorporating the latest context information to produce a new candidate. By utilizing feedback from multiple commit message quality evaluators, our method allows the LLM to generate higher-quality commit messages through an ongoing optimization process iteratively.

\section{RQ1: Categorization of Missing Software Context Information}
\label{sec:rq1}

% for space limit I 'd like to delete this sentence below

Li et al. \cite{li2024only} used GPT-4 in their OMG approach to generate commit messages, achieving state-of-the-art results in CMG. 
However, since commit messages are a key communication channel for discussing diverse maintenance activities across various software contexts \cite{tian2022makes, li2023commit}, we argue that software developers have considerable freedom to consider any information sources that originate from various places in the software repository or even interpersonal discussions and hence OMG overlooks many crucial information. For example, as shown in Figure \ref{figure:im}, understanding the definition and usage of the invoked method \textit{getComputer} is essential to generate a commit message that aligns with human-written ones. OMG generated a message as \textit{``Fix: Exclude ``master'' from testGetComputerView() test. In the testGetComputerView() method in the ComputerClientLiveTest.java file, a condition has been added to exclude the ``master'' from the test. This change ensures that the test only validates the display name of each computer in the view, excluding the ``master''.''}, which reflects that it failed to cover the information of this invoked method in the generated commit message. Instead, it only captures the textual content of the source code at a surface level (``\textit{...exclude the ``master" from the test.}'', ``\textit{only validates the display name of each computer in the view, excluding the ``master".}''). In this study, we investigate the software contexts that are considered by human software developers when writing commit messages but are missed by OMG. We present the methodology in (Section \ref{sec:rq1_method}) and the results in (Section \ref{sec:res_rq1}). 

%In this section, we present the methodology (Section \ref{sec:rq1_method}) and results (Section \ref{sec:res_rq1}) for our \textbf{RQ1: What software context information do developers consider when writing commit messages that has been missed by the CMG techniques?}

\subsection{Methodology to Answer RQ1}
\label{sec:rq1_method}

We began by having two authors independently review the code changes, human-written commit messages, and OMG-generated messages using a card sorting process \cite{zimmermann2016card}. The dataset comprised 381 commits across 32 Apache projects that were examined by Li et al. \cite{li2024only} (a detailed description of the dataset is in Section \ref{sec:dataset}). The authors assessed various software contexts related to the code changes, including code files, comments in issue reports/pull requests, and commit history. Each author first analyzed the code changes and human-written commit messages to identify the software contexts likely considered by the developers. Then, they verified whether the developer indeed considered these contexts when writing the message by closely examining the drafted contexts. For example, as shown in Figure \ref{figure:im}, the authors might draft the information of the invoked method \textit{getComputer} as a considered context by the developer based on the code changes and the human-written message. To verify the context, they manually checked the body of this method to examine whether the awareness of this invoked method would help write this message. Similarly, the authors also identified and verified the contexts considered in OMG-generated messages. Contexts considered by humans but missed by OMG were retained and categorized into codes. The authors then discussed these codes, organizing them into high-level categories, which were further refined into meaningful themes. The themes are detailed in Section \ref{sec:res_rq1}.

\subsection{Answer to RQ1: Software Context Missed by CMG}
\label{sec:res_rq1}

\subsubsection{Context Themes and Frequency}
\label{sec:themes}
We identified 8 themes of software context that humans considered, but OMG missed. Specifically, 46.5\% (177/381) of OMG-generated messages failed to include some software contexts recognized by humans (for dataset details, see Section \ref{sec:setup}). Notably, a single commit can exhibit multiple contexts missed by OMG but considered by humans. The themes are described as follows:

%OLD: We identify 8 themes of software context that is considered by humans but missed by OMG. In total, 46.5\% (177/381) of the OMG-generated messages (the dataset is described in Section \ref{sec:setup}) missed some software context that is considered by humans. It's worth noting that multiple themes of context can be missed by OMG but considered by humans for a single commit. We describe these themes as follows:

\noindent\textbf{1. Ad-hoc Software Maintenance Goals} (23.1\%, 88/381): 
% DF: I do not understand this class name:  "Software Maintenance Activity" is defined very general in the paragraph, covering almost any code change. So I guess this is about "Ad-hoc" and "Ad-hoc" means without an issue being referenced. So this class covers all commits where no corresponding issue exists? 
% DF: The amount of issues this class has suggests it is a kind of misc class. Looking at the first 10 examples in this class, I agree that is is very hard to write good commit messages. I did not find any pattern to categorize them. I think here the class name and description should be updated, or the set of examples even partitioned into different classes. From the first 10 examples, I think the class is correctly judged as "impossible".
This theme involves developers describing code changes with reference to a specific maintenance goal, such as functional correction, new feature addition, or non-functional improvement, without citing an issue report or pull request. Instead, the commit message itself conveys the maintenance goal. For example, a message like \textit{``change required after plexus update"} \cite{adhoc} indicates that the changes ensure compatibility with an updated third-party dependency, with the maintenance goal serving as context for explaining the code changes.

%OLD: In this case, the developers aim to describe the code changes from the perspective of a specific software maintenance goal that the code changes were committed to achieve. The goal can fall into any maintenance activity category such as functional correction, new feature addition, and non-functional improvement. However, there is no referenced issue report/pull request -- all maintenance goals are described textually in the commit message. For example, \textit{``change required after plexus update"} \cite{adhoc} indicates the committed code changes are for making the code compatible with an update of a third-party dependency. The developers adopted this maintenance goal as a context to describe the motivation behind the code changes. 

% https://github.com/apache/archiva/commit/ce29219ee6409b111ce8f8191b1792c2b65a04a8
% \noindent\textit{Example}: The example in Figure \ref{} shows that the developers describe a software maintenance activity in the commit message -- the committed code changes are for making the code compatible with an update of a third-party dependency. Since the dependency (\textit{``plexus"}) is not contained in the code change representation (i.e., git diff), this maintenance activity is adopted as a context to describe the motivation behind the code changes. 

\begin{figure}[htp]
\label{figure:examples}

\centering
\begin{subfigure}{\textwidth}
\centering
\includegraphics[width=0.55\textwidth]{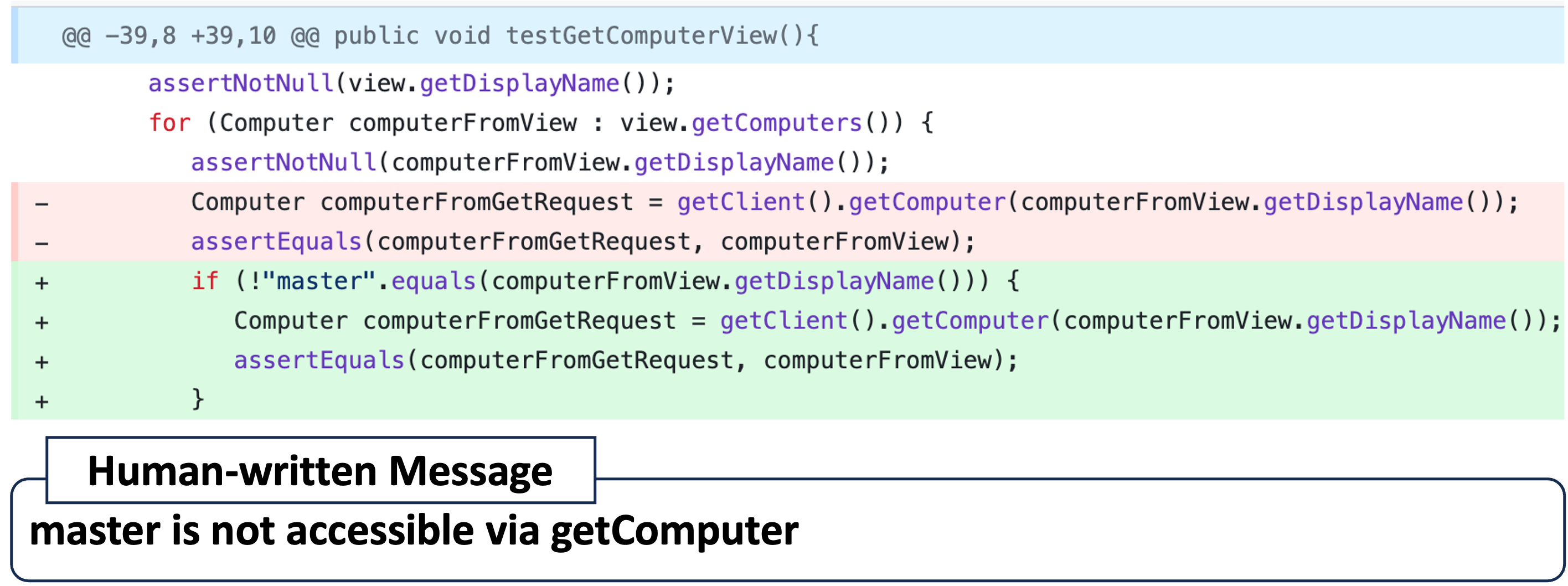}
\caption{Example of Invoked Methods}
\label{figure:im}
\end{subfigure}

\bigskip

\centering
\begin{subfigure}{\textwidth}
\centering
\includegraphics[width=0.55\textwidth]{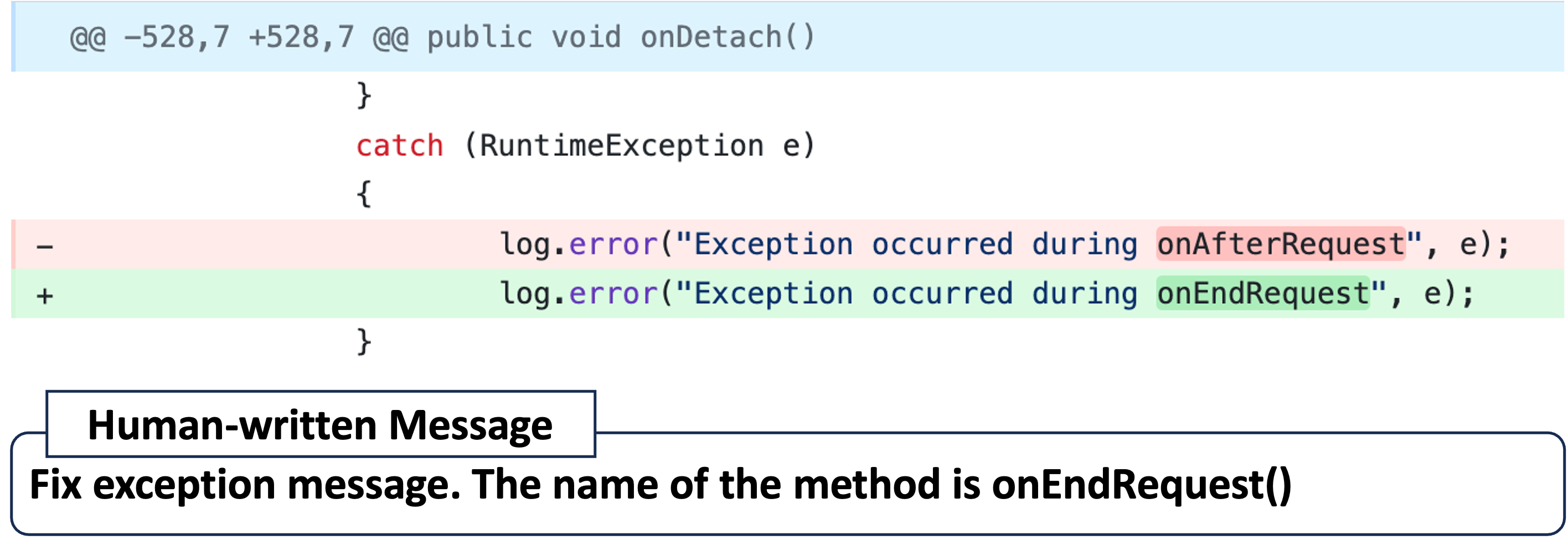}
\caption{Example of Syntactically Related Code Blocks}
\label{figure:cb}
\end{subfigure}

\caption{Example Commits}

\end{figure}

\noindent\textbf{2. Invoked Methods} (11.5\%, 44/381): 
This theme occurs when developers modify method calls or make changes related to method calls, leading them to describe the ``What''/``Why'' based on the invoked methods. For example, as shown in Figure \ref{figure:im} \cite{invoked}, developers used the invoked method to explain the motivation behind the changes. In this case, the addition of the \textit{if} statement is due to the fact that \textit{``master"} is inaccessible via \textit{getComputer}. Understanding the method \textit{getComputer} was essential for crafting this commit message.

%OLD: The developers rely on their knowledge about the invoked methods within the code changes to write the commit message. It happens when the developers edit method calls or the committed code changes are related to some method calls in the code, which leads to them describing the ``What''/``Why'' information based on the knowledge of the invoked methods. Figure \ref{figure:im} \cite{invoked} shows an example where developers relied on an invoked method to explain the motivation behind the code changes (i.e., ``Why''). The addition of the \textit{if} statement stems from the fact that \textit{``master"} is not accessible via \textit{getComputer}. Only by possessing the knowledge of the invoked method \textit{getComputer} could the developer write such a commit message.

% \begin{figure*}[t!]
%     \centering
%     \begin{subfigure}[t]{0.5\textwidth}
%         \centering
%         \includegraphics[width=\textwidth]{figures/invokedmethod3.png}
%         \caption{Lorem ipsum}
%     \end{subfigure}%
%     ~ 
%     \begin{subfigure}[t]{0.5\textwidth}
%         \centering
%         \includegraphics[width=\textwidth]{figures/codeblock4.png}
%         \caption{Lorem ipsum, }
%     \end{subfigure}
%     \caption{Caption place holder}
% \end{figure*}

% \noindent\textit{Example}: Figure \ref{} shows an example where developers rely on an invoked method to explain the motivation behind the code changes (i.e., ``Why"). The addition of the \textit{if} statement stems from the fact that \textit{``master"} is not accessible via \textit{getComputer} where the knowledge of the invoked method \textit{getComputer} would be necessary to write this message. 

\noindent\textbf{3. Project Requirements} (11.3\%, 43/381): 
% DF: checked first 15 examples: 60% of project requirements are actually software engineering requirements, aka best practices (Don't swallow IOException, One should always run unit tests before checking files in, Fix incorrect finalize override, make it consistent., register the typeInfos correctly, make sure streams get closed, Stop using deprecated methods, use additional overloads for the toArray method, remove unused body). The LLM should be able to generate good commit messages if prompted like "If there is a software development best practice that motivates this change, write a commit message very shortly describing this best practice. Otherwise say 'No best practice motivates this change'".
Developers often implicitly rely on project requirements or recommended practices to explain code changes. Phrases like ``should,'' ``do not,'' or ``allow'' suggest that a project requirement or convention influenced the decision, though developers did not reference specific requirements directly. Instead, they use certain words to imply the necessity of the change. For example, \textit{``Don't try to config mdb destination if we aren't auto creating resources.''} \cite{projreq} suggests that the change was made to correct a violation of a project requirement, even though the specific requirement is not explicitly mentioned.

\noindent\textbf{4. Variable Data Types} (7.1\%, 27/381): Understanding the class or data type of an edited variable helps explain the code changes in the commit message. For example, in commit \cite{datatype}, developers added several methods, each returning a variable. The message \textit{``Add getters for private ivars''} highlights that the variables' data types and access modifiers were considered.

%OLD: The understanding of the class that the edited variable is an instance of or its data type contributes to the explanation of the code changes in the commit message. For example, the developers implemented several new methods in commit \cite{datatype}. Each of these functions returns a variable. The message was written as \textit{``Add getters for private ivars"}, which indicates the variables' data types including their access modifiers in the class are used to describe the added methods.

% https://github.com/apache/logging-log4j2/commit/856607faf268136fc6f2e27cf3716a5e16a0f48d
% \noindent\textit{Example}: Figure \ref{} presents an example where the data types of these returned variables including their access modifiers are needed to explain the reason for the added getter methods. 

\noindent\textbf{5. Related Code Changes in History} (4.5\%, 17/381): 
% DF: 70% estimated success for searching in commit history with fixed rules like: (a) pick the last i commits from the same author (b) pick the last j commits (c) pick the last k commits modifying a superset of the files being modified in this commit (d) pick the last l commits modifying a superset of LOCs being modified in this commit. 80% estimated success when using AI not only to process the set of retrieved former commits, but also decide on which commits to retrieve based on the code change and history commits retrieved so far (explain the AI how to use git command line and let it loose).
Code changes from the project's history are often referenced when writing commit messages for a given commit. These historical changes may relate to the current commit in various ways, such as fixing defect-introducing changes, enhancing earlier features, or repurposing previous modifications. For example, \textit{``Re-enabling test as it seems to have been fixed''} \cite{commithist} indicates that re-enabling the test case in the current commit is linked to a past commit. Here, the previous changes serve as context for composing the message.

%OLD: The code changes in the history of the software project are referenced as a context to write the commit message for the given commit. These code changes in the history are usually related to the code changes in the current commit in various ways, such as fixing defect-introducing changes, enhancing previously introduced features, or using previously introduced code changes for other purposes. For example, \textit{``Re-enabling test as it seems to have been fixed."} \cite{commithist} shows that re-enabling the test case in the current commit is closely related to a commit in the past (\textit{``...it seems to have been fixed"}). The past code changes were used as a context to compose the commit message.

% https://github.com/apache/jmeter/commit/ffebf3c0589a4793969b0e5b3d21bacfb03c2e49
% \noindent\textit{Example}: In this example shown in Figure \ref{}, re-enabling the test case in the current commit is closely related to a software maintenance activity in the past (\textit{``...it seems to have been fixed"}, \textit{``...Former-commit-id: 399d67f"}). This past software maintenance activity is used as a context to compose the commit message.

\noindent\textbf{6. Syntactically Related Code Blocks} (3.4\%, 13/381): The surrounding source code is often considered when writing commit messages. However, the limited length of the git diff—which represents the code changes -- may not provide sufficient context for developers or CMG approaches to fully explain or summarize the changes. Additional context, such as the entire enclosing statement block, is often needed but may not be included in the git diff. OMG missed this information because its summaries of the enclosing method or class captured only high-level functionality, leaving out important detailed source code information. Hence this theme focuses on the textual source code near the changed lines, as opposed to the broader summaries of the method or class considered by OMG. For example, accurately describing the changes in Figure \ref{figure:cb} \cite{codeblock} (i.e., \textit{onEndRequest}) requires access to the full enclosing \textit{try-catch} block to identify the executed method in the \textit{try} section. Relying solely on the git diff would miss this crucial context, as reflected in the associated human-written commit message.

\noindent\textbf{7. Personal Mistakes} (2.9\%, 11/381): 
% DF: 1 of 11 ("Reverting unintentional change") could be fixed via the same approach as "Related Comits in the History"). Actually, I think I would categorize that commit into "Related Commits In the History. Most of the 11 examples are really hard to generate commit messages for. Furthermore, it is a class with likely high variance, making it even harder. 
The code changes are made to resolve some personal mistakes developers have made in the past. These mistakes can include copy-paste errors, accidents due to negligence, and typos. %This context information stems from the awareness of the mistakes that the developers made by themselves in the past or the communication between developers that leads to the awareness of mistakes made by others. 
For example, \textit{``fix typo: wrong if guard variable''} \cite{mistake} shows the reason for the \textit{wrong if guard variable} is the fact that some developer made a typo. In addition, some developers may simply write commit messages as \textit{``Remove author tag. Thanks Sylvain for pointing at this, this happens when you copy paste and don't think about what you're doing.''} \cite{personalmistake1} or \textit{``Remove getFilter method inadvertantly left in''} \cite{personalmistake2}.

% https://github.com/apache/cassandra/commit/eaced9a541d09d55973b6f88d720e16ac948a559
%\noindent\textit{Example}: In the commit shown in Figure \ref{}, the reason for the \textit{wrong if guard variable} is the fact that some developer made a typo. 

\noindent\textbf{8. Compile/run-time Information} (2.4\%, 9/381): When developers commit code changes to fix compilation or run-time errors, they often specify this in the commit message. For example, \textit{``Resolve trivial compilation error after previous merge''} \cite{compile} indicates that the changes address a compile-time error, reflecting the developers' understanding of the software’s compile-time behavior.

\section{Experimental Settings}
\label{sec:setup}

In this section, we provide the details about the experimental settings for answering our RQ2 (Section \ref{sec:rq2}) and RQ3 (Section \ref{sec:rq3}).

\subsection{Baselines}
\label{sec:baselines}

We selected three state-of-the-art techniques as baselines. The first was OMG, an LLM-based CMG approach \cite{li2024only} utilizing ReAct prompting. The second was CMC \cite{eliseeva2023commit} chosen because both CMC and our approach are guided by human input. Lastly, we used human-written commit messages as a baseline to assess whether our technique enhances their quality.

\subsection{Evaluation Metrics}
\label{sec:metrics}

\subsubsection{Commit Message Evaluation Metrics}
\label{sec:hm_metrics}
Similar to OMG \cite{li2024only} we used the following four metrics for a comprehensive commit message quality evaluation. The  metrics are: 

%Since Li et al. \cite{li2024only} proposed four evaluation metrics by merging developers' expectations for commit message quality and human evaluation metrics used by CMG approaches in literature \cite{dong2022fira,shi2022race,he2023come,wang2023delving}, we directly adopted these metrics for a comprehensive commit message quality evaluation. The four metrics are:

\noindent\textbf{Rationality}: it assesses whether the commit message provides a logical explanation for the code changes (``Why'' information), and provides the commit type.

\noindent\textbf{Comprehensiveness}: it assesses whether the commit message contains a summary of the code changes (``What'' information), and covers all changed files.

\noindent\textbf{Conciseness}: it evaluates whether the commit message conveys information succinctly, ensuring readability and quick comprehension.

\noindent\textbf{Expressiveness}: it reflects whether the commit message is grammatically correct and fluent.

% In this study, we obtained the evaluation results in these metrics from both our automated evaluators (Section \ref{sec:obj}) and humans. 

% I need to put these somewhere...
%Li et al. \cite{li2024only} asked humans to score the commit messages in terms of these metrics on a 5-point Likert scale (0 for poor, 1 for marginal, 2 for acceptable, 3 for good, and 4 for excellent). The results showed that 63.5\% of the OMG-generated messages were labeled as 4 across all the four metrics, while we found OMG could miss necessary software context information that is considered by software developers (Section \ref{}). We posit that the 5-point Likert scale by Li et al. would be insufficient in our human evaluation since our approach might improve the generated messages with a score of 4. To resolve this issue and better evaluate the messages, we asked four researchers with at least five years of Java development experience to independently rank the messages generated by OMG, OMG+, and CMG-AllPrompt, completed by LLM-prompted CMC, and optimized by our approach, based on the quality evaluated by the four evaluation metrics. They were also asked to answer whether the generated commit messages were better than the human-written ones. It's worth noting that the researchers were not aware of where the messages had been generated from except the human-written ones that were needed for comparison. During this process, the researchers were allowed to examine all software contexts associated with the code changes to evaluate the quality of the commit messages. 

\subsubsection{Traditional Evaluation Metrics}
\label{sec:bleu}
%\hl{Jiawei: Since we are optimizing based on the human-written messages, I don't think we should still use these NLP metrics? Also, researchers showed these metrics are not good in CMG and LLM-generated messages are different from human-written ones. I don't think we need this evaluation.}

We also adopted traditional metrics that are widely used in previous CMG works %\cite{jiang2017automatically,liu2020atom,dong2022fira,he2023come,li2024only} to evaluate the quality of the generated commit message, 
including BLEU \cite{papineni2002bleu}, METEOR \cite{banerjee2005meteor}, and ROUGE-L \cite{lin2004automatic}. The human-written messages are used as references. 

%\hl{Jiawei: Can we mention the reason why we don't use these metrics in this section? It shows: yes we know these metrics but they are just not applicable here.}

\subsection{Human Evaluation}
\label{sec:humaneval}

We also conducted a human evaluation where two researchers, each with over five years of Java development experience, independently evaluated 76 commits from the testing split of the \textit{LLM-based Quality Evaluator} (Section \ref{sec:obj}). They ranked eight commit messages, including the CMO-optimized message, the baseline messages (human-written, OMG, CMC), and messages from CMO variants in the ablation study (Section \ref{sec:rq3}), based on four metrics (Section \ref{sec:hm_metrics}). To ensure fairness, the evaluators were blinded to the message sources, including the human-written ones, but had access to all relevant software contexts associated with the code changes.

As part of the human evaluation, we also surveyed 22 Apache OSS developers to evaluate commit messages, adhering to IRB-approved protocols. Developers received surveys containing git diffs and four corresponding messages (human-written, OMG-generated, CMC-completed, and CMO-optimized) for each diff. They ranked the messages using four metrics (Section \ref{sec:hm_metrics}). To avoid overwhelming them, instead of sending 76 diffs like the researcher evaluation, we randomly sampled 10 out of 76 diffs and their associated messages, as large workloads can lower survey completion rates \cite{smith2013improving}. Like the researcher evaluation, developers were unaware of the message's origins but could review relevant software context.

%Moreover, to further investigate whether CMO outperforms the existing state-of-the-art baseline techniques and optimizes the human-written messages (Section \ref{sec:baselines}) from the perspective of real-world developers, we asked 22 Apache Open-Source Software (OSS) developers to evaluate the messages. Specifically, we sent surveys (following university-approved Institutional Review Boards (IRB) protocol) containing git diffs and the four messages to them (the human-written, OMG-generated, CMC-completed, CMO-optimized messages), where they were asked to rank the messages for each git diff in terms of the four metrics. Since it's manually demanding for the developers to check 76 git diffs with 304 messages, we randomly sampled ten diffs and their associated four messages for them to rank to ensure they would finish the surveys (large workload has been shown to decrease the completion rate for survey participates \cite{smith2013improving}). Similar to the evaluation by the researchers, these developers were not aware of the origins of the provided messages, but were allowed to check any software contexts associated with the code changes.

\subsection{Dataset}
\label{sec:dataset}

We used the dataset collected by OMG \cite{li2024only}, which includes 381 commits from 32 Apache Java projects. Each commit's code diff is paired with three commit messages: one human-written, one generated by OMG, and one by FIRA \cite{dong2022fira}, a CMG technique that OMG outperformed. All commit messages have been manually evaluated and scored using the evaluation metrics detailed in Section \ref{sec:hm_metrics}, with scores assigned on a 5-point Likert scale (0 for poor, 1 for marginal, 2 for acceptable, 3 for good, and 4 for excellent).

%OLD: We selected the dataset collected by OMG \cite{li2024only} in this study (the detailed statistics of the dataset is in \cite{li2024only}). The dataset contains 381 commits from 32 Apache projects in Java. The git diff of each commit has three associated commit messages that are written by humans, generated by OMG and FIRA \cite{dong2022fira} (a CMG technique that OMG compared with to show its superior performance). All commit messages in this dataset have been manually evaluated and labeled in scores based on the evaluation metrics in Section \ref{sec:hm_metrics}. The scores are on a 5-point Likert scale (0 for poor, 1 for marginal, 2 for acceptable, 3 for good, and 4 for excellent). 

\section{Commit Message Optimization (CMO)}
\label{sec:cmo}

%\subsection{Motivation}
\subsection{Tools for Automated Software Context Retrieval}
\label{sec:tools}

To support the LLMs' reasoning and generation of commit messages that are more factual and grounded on the code base, similar to OMG, we try to implement tools to automatically retrieve the identified software contexts in Section \ref{sec:themes} that were missed by OMG.  These tools aim to capture the human-considered contexts, further enhancing the quality of LLM-generated commit messages.

%OLD: To support the LLMs' reasoning and generation of commit messages that are more factual and grounded on the code base, OMG utilized a tool module to automatically retrieve several types of software context information such as the title of issue report/pull request and the summary of method/class enclosing the code changes to feed into the LLMs. Similarly, we try to implement tools to automatically retrieve the identified software contexts in Section \ref{sec:themes} that were missed by OMG, aiming to help the LLMs generate commit messages that cover the necessary human-considered contexts to further improve its quality.  

For extracting \textit{Invoked Methods}, \textit{Variable Data Types}, and \textit{Syntactically Related Code Blocks}, we utilized JavaParser \cite{javaparser}, a widely-used tool for parsing Java program elements \cite{wang2020empirical,li2024only}. JavaParser helped retrieve information about invoked methods, variables, and statement blocks related to code changes. Specifically, we identified the data types and class information of variables referenced in the modified code lines. Additionally, we extracted the smallest statement blocks that enclose the changed code, as these often provide sufficient context for composing human-written commit messages for \textit{Syntactically Related Code Blocks} (Section \ref{sec:themes}). Statement blocks, delimited by curly brackets, included all types supported by JavaParser.
For \textit{Invoked Methods}, we extracted the method bodies called within the modified code and summarized them using a state-of-the-art method-level code summarization technique \cite{geng2024large}, also used by OMG, to manage the limited input context length of LLMs \cite{vaswani2017attention}. This study focused on extracting invoked methods and variables' data types defined within the project's source files or as part of the Java language. Retrieving methods and data types from third-party libraries remains a future task.

However, we did not implement automated tools to retrieve information for other software context themes in Section \ref{sec:themes} due to the impracticality of automatically retrieving them and the potential noise brought by this process. To support our decision and assess the feasibility of implementing the retrieval tools for these themes, we also consulted with two senior Java software engineers with more than ten years of Java software development and commit message writing experience.

For \textit{Ad-hoc Software Maintenance Goals}, code changes are often implemented to achieve specific maintenance objectives (``ad-hoc’’ goals). These goals may be documented in related software artifacts, such as issue reports, pull requests, or discussions. However, accurately identifying the maintenance goal behind a code change becomes nearly impossible to automate without explicit references to these artifacts. One possible solution is to use an issue-commit link traceability recovery technique. However, their current performance only achieves a probability of 0.1 to 0.5 for correctly predicting issue-commit links \cite{zhang2023ealink}, which could introduce significant noise into LLM-generated commit messages. Improving these techniques is beyond the scope of this study.

Another approach could be to compile and run the code before and after the change to detect performance improvements or bug fixes. This process would require a variety of analyzers and tests, including reproducing issues by compiling and running the software in both versions. To achieve this goal, we reset the software repositories to their pre- and post-change versions, examined the project structures, set up environments, and manually resolved dependencies. Despite these efforts, we could only compile and run 33\% of the versions successfully. The failures may have been due to actual bugs being fixed, errors in understanding project dependencies, or other unrelated issues. Hence, to avoid introducing noise, we did not develop tools for retrieving compile/run-time information to help LLMs identify maintenance goals, nor did we implement retrieval tools for \textit{Compile/run-time Information}.

For \textit{Project Requirements}, insufficient detail about the specific project requirements that developers referenced when making code changes is available. The manual effort needed to create traceability links between source code and requirements is significant, and often absent or incomplete in most software projects \cite{hey2021improving}. Moreover, existing approaches for recovering these traceability links generally perform poorly, with F-1 scores below 0.5~\cite{zhao2017improved,chen2019enhancing,moran2020improving,hey2021improving}, which could introduce noise if used in our context. As improving these techniques is beyond the scope of this work, we did not explicitly develop tools to extract potential linked requirements from the code changes.

For \textit{Related Code Changes in History}, we explored heuristics to retrieve relevant commits automatically but found diverse patterns in how developers refer to past commits. Specifically, developers considered (1) the most recent commit that changed the same lines as the current commit (4/17, 23.5\%); (2) the most recent commit that changed the same functions, though not necessarily the same lines (3/17, 17.6\%); (3) a specific commit in the history affecting the same lines or functions (4/17, 23.5\%); and (4) a previous commit impacting code units that the current code depends on or demonstrates motivation for the current changes (6/17, 35.3\%). Each rationale depends on the specific code changes and context, making it difficult to determine which historical commit is relevant. Implementing a retrieval tool based on any one of these patterns would likely introduce noise to the LLM-generated commit messages. After consulting with developers, we chose not to build such a tool. Similarly, automating the extraction of \textit{Personal Mistakes}--a context residing solely in developers' minds during code changes–is impractical or impossible.

\subsection{Motivation for Developing CMO}
Since many of the contexts frequently considered by developers are missed by CMG techniques and cannot be reliably extracted with automated tools, human-written commit messages could help address this gap. These messages naturally include software contexts that are difficult or impractical to extract automatically. Incorporating human-written messages as a guide in CMG could reduce the need for substantial effort in retrieving such contexts. However, the value of human guidance has been overlooked in CMG techniques, despite its successful application in automating other software engineering tasks, such as program repair and assertion generation~\cite{bohme2020human,zamprogno2022dynamic}.

%OLD: However, without the access to these contexts that are frequently considered by developers, CMG techniques can miss certain details and generate sub-optimal commit messages. Since the human-written commit messages would include these impractical-to-extract software contexts, incorporating these messages as a guide in CMG would ideally spare the substantial efforts and overcome the impracticality for the retrieval of such contexts. However, the guidance from humans has been overlooked by CMG techniques, which has been adopted in the automation of other SE tasks such as program repair and assertion generation \cite{bohme2020human,zamprogno2022dynamic}. 

To incorporate human guidance and simplify the CMG task for developers, Eliseeva et al. \cite{eliseeva2023commit} introduced the CMC approach. In CMC, a git diff and the beginning of a human-written commit message are provided as input to CMG techniques, which only complete the remaining part of the message rather than generating it from scratch. This method allows for the creation of commit messages that are more semantically aligned with human-written ones. However, the authors did not comprehensively evaluate the quality of these completed messages, nor did they account for the essential software contexts needed to generate high-quality commit messages \cite{tian2022makes,li2024only}.

%OLD:Aiming to bring in human guidance and make the task of CMG easier and more practical to developers, Eliseeva et al. \cite{eliseeva2023commit} proposed CMC. In CMC, a git diff along with a prefix of its human-written commit message is input into various CMG techniques. These techniques only need to complete the message by generating the remaining tokens instead of the whole commit message from scratch. Since CMC spares CMG techniques from generating the messages from the beginning, it can generate messages that are more semantically similar to human-written ones. However, the authors did not evaluate the quality of the completed messages comprehensively nor incorporate the associated software contexts that are essential to the generation of high-quality commit messages \cite{tian2022makes,li2023commit,li2024only}. 

In this study, we advance beyond CMC by introducing \textit{Commit Message Optimization} (CMO), aiming to enhance the quality of generated commit messages. Unlike CMC, which completes a given prefix of a commit message, CMO optimizes an existing human-written commit message to achieve the highest quality. This approach leverages both the human-written message and other frequently considered contexts by developers. By using the full commit message as input, CMO minimizes information loss and preserves the valuable human-considered contexts. The following sections detail the design and implementation of CMO.

%OLD: In this study, we take a further step to move from CMC to \textit{Commit Message Optimization} (CMO) with the goal of improving the quality of the generated messages by incorporating human written message in the process: instead of generating a commit message from scratch (CMG) or completing a prefix (CMC), CMO optimizes an existing human-written commit message to be of the highest quality by considering both the information in the human-written message and other frequently considered contexts by developers. We posit that using the whole commit message as the input would avoid the information loss incurred by only using a prefix so that the human-considered contexts would be preserved. In the following sections, we describe the design and implementation of CMO.

%Specifically, CMO takes a git diff and its associated human-written commit message as input at the beginning. It iteratively and continuously updates the human-written message by considering different types of software contexts that are frequently considered by developers (Section \ref{sec:res_rq1}). The goal of the updates is to optimize our objective function, where we build several automated commit message quality evaluators and use the evaluation result as the objective function. Finally, a search-based algorithm explores the generated/updated message candidates and identifies the best one as the optimized message for the input. In the following sections, we describe the design and implementation of CMO in detail.

\subsection{Objective Function}
\label{sec:obj}
In this study, we aim to enhance the quality of generated commit messages using four key metrics detailed in Section \ref{sec:hm_metrics}. These metrics represent widely accepted human evaluation criteria from the CMG literature and align with software developers' expectations for commit message quality \cite{li2024only}. To construct an objective function that maximizes these metrics, we employ an end-to-end automated approach for scoring the messages based on their quality. We have designed two evaluators for this purpose: the \textit{LLM-based Quality Evaluator} and the \textit{Retrieval-based Quality Evaluator}. The \textit{LLM-based Quality Evaluator} uses a fine-tuned LLM to assess messages against the four metrics. In contrast, the \textit{Retrieval-based Quality Evaluator} evaluates the quality of generated messages by comparing them to high-quality human-written messages associated with similar git diffs. This evaluator also integrates additional contextual information that might influence human judgment. To provide a comprehensive assessment, we combine the results from both evaluators to produce a single score for each of the four metrics, which serves as the objective function for our optimization process. We detail the evaluators and the combined evaluation score calculation mechanism below.

\noindent\textbf{1) LLM-based Quality Evaluator}: Recent research shows that LLMs can be robust evaluators across various generation tasks through prompting or fine-tuning \cite{kim2023prometheus,yao2024tree,gao2024llm}. In this study, we fine-tuned GPT-3.5-Turbo \cite{gpt35} (as GPT-4 is not yet available for fine-tuning at the time of analysis) to automatically evaluate commit messages based on the four metrics in Section \ref{sec:hm_metrics}. Using the dataset in Section\ref{sec:dataset}, we split it into 80\% training (305 commits) and 20\% testing (76 commits). Since each commit in the dataset is linked to three different commit messages (human, FIRA, and OMG), the training set contained 915 messages, and the testing set had 228, each labeled with four human scores for the metrics. 
We treated this as a multi-class classification task, with Likert scores (0–4) as class labels, fine-tuning GPT-3.5-Turbo to predict scores based on git diffs and commit messages. We fine-tuned four models, one for each metric. To address class imbalance (e.g., only 8.6\% of messages were labeled with a score of 3 in \textit{Rationality}), we used random oversampling \cite{kotsiantis2006handling} via \textit{imblearn} \cite{imblearn} to balance the training data across all classes. We present the performance of the classifiers on the testing split in Table \ref{tab:llmeval}. It's worth noting that we calculated precision, recall, and F1-score for each class and then obtained their average weighted by the number of true instances for each class\cite{clsmetrics}.

\begin{table}[]
\footnotesize
\setlength{\tabcolsep}{2pt}
\caption{The Performance of LLM-based Quality Evaluators}
\label{tab:llmeval}
\begin{tabular}{l|cccc}
\hline
                  & \textbf{Accuracy} & \textbf{Precision} & \textbf{Recall} & \textbf{F1} \\ \hline
Rationality       & 0.719             & 0.653              & 0.719           & 0.684       \\ \hline
Comprehensiveness & 0.719             & 0.695              & 0.719           & 0.660       \\ \hline
Conciseness       & 0.895             & 0.801              & 0.895           & 0.845       \\ \hline
Expressiveness    & 0.912             & 0.849              & 0.912           & 0.875       \\ \hline
\end{tabular}
\end{table}

\begin{figure*}[t]
    \centering
    \includegraphics[width=0.4\textwidth]{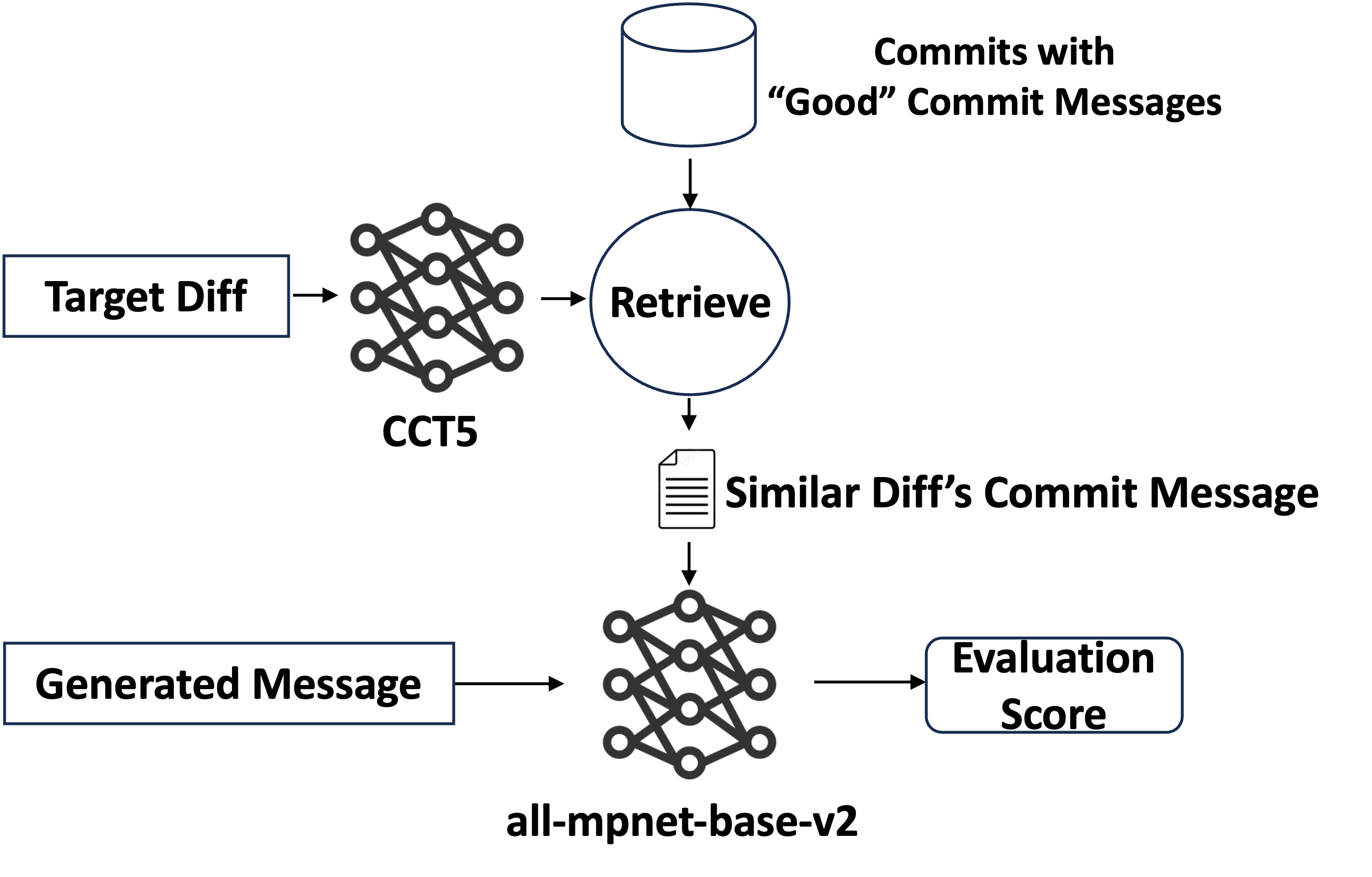}
    \caption{Overview of Retrieval-based Quality Evaluator}
    \label{fig:retrieval-eval}
\end{figure*}

\noindent\textbf{2) Retrieval-based Quality Evaluator}: 
%Since a human-written commit message can be of low quality without any ``What''/``Why'' information \cite{tian2022makes, li2023commit}, such a message may bring noise or mislead the optimization to generate a sub-optimal message. However, human's guidance is still preferred as it's not always practical to automatically retrieve all human-considered contexts missed by CMG techniques (Section \ref{sec:res_rq1}).In order to still exploit the contexts that are considered by humans and mitigate the potential negative effect brought by low-quality human-written messages, we create this \textit{Retrieval-based Quality Evaluator} as another major component of our objective function, 
This evaluator gauges the quality of generated commit messages by measuring their semantic similarity to high-quality human-written messages that cover both the ``What'' and ``Why'' aspects. As depicted in Figure \ref{fig:retrieval-eval}, the process starts by retrieving git diffs from a data corpus that are semantically similar to the target diff being optimized. The evaluator then compares the generated commit message with a high-quality human-written message, using semantic similarity as the evaluation score. This approach indirectly incorporates human guidance, as it benchmarks the generated message against what a skilled human might write for similar code changes.

%This evaluator evaluates the generated commit messages based on their semantic similarity to the high-quality human-written messages with both ``What'' and ``Why'' information. Figure \ref{fig:retrieval-eval} shows an overview of this evaluator. The evaluator first retrieves the git diff that is semantically similar to the target diff that we optimize the commit message for from a data corpus. Then, the generated commit message is compared with the human-written message of the similar git diff where their semantic similarity is used as the final evaluation score. %Since the decision is based on comparing with a good quality human written message, the human's guidance is impliclty incorporated in the message generation process. 

To construct the data corpus for retrieving similar git diffs, we collected commits (i.e., git diffs and human-written messages) from 32 Apache projects utilized by OMG \cite{li2024only} and other studies \cite{mannan2020relationship,li2023commit}. We filtered the commits to include only those with both ``What'' and ``Why'' information, deemed ``Good'' commit messages \cite{tian2022makes}, using a state-of-the-art classifier from Li et al. \cite{li2023commit} that classifies whether a commit message contains ``What'' or ``Why'' information.

%Specifically, to construct the data corpus for the retrieval of similar git diffs, we collected the commits (i.e., git diffs and human-written messages) from the 32 Apache projects used by OMG \cite{li2024only} and other previous works \cite{mannan2020relationship, li2023commit}. To ensure the quality of the human-written messages in the collected commits, we only retained those commits whose messages have both ``What'' and ``Why'' information (``Good'' commit messages \cite{tian2022makes}), and eliminated others from the data corpus. To do so, we used a state-of-the-art classifier by Li et al. \cite{li2023commit} that classifies whether a commit message contains ``What'' or ``Why'' information. 

For semantic similarity retrieval, we employed CCT5 \cite{lin2023cct5}, a pre-trained language model capable of capturing the semantic essence of code changes. We used the vectorized representation of the special token [CLS] from CCT5's last encoder layer to represent each git diff. We then applied cosine similarity to these vectorized representations to find git diffs similar to the target diff. To assess the semantic similarity of commit messages, we used \textit{all-mpnet-base-v2} \cite{mpnet}, a pre-trained natural language model noted for its performance on the Sentence Transformer Leader Board \cite{sentencetransformer}. We converted the commit messages into vectorized representations and calculated cosine similarity between the generated and human-written messages from the most semantically similar git diff \cite{liu2018neural,gao2023makes,geng2024large}. This cosine similarity score reflects how closely the generated message aligns with high-quality human-written messages and was used as the evaluation score. To mitigate potential bias from relying solely on the commit message of the most similar git diff, we experimented with multiple similar git diffs. Specifically, we retrieved the top 1, 5, 10, and 20 git diffs that were most similar to the target diff. The average of the cosine similarity scores between the generated message and the human-written messages for these diffs was then used as the final evaluation score.

\noindent\textbf{3. Combined Evaluation Score}: In this study, the goal is to optimize human-written commit messages to maximize their quality measured by the four evaluation metrics in Section\ref{sec:hm_metrics}. To provide a comprehensive assessment, we combine the results from both evaluators
to produce a single score for each of the four metrics, which serves as the objective function. 

To normalize the evaluation score of the \textit{Retrieval-based Quality Evaluator} (denoted as \textit{``Sim Score''}), which ranges from 0 to 1, we scaled it to a range of 0 to 4 by multiplying the score by 4. This allowed us to directly compare it with the \textit{LLM-based Quality Evaluator} score (denoted as \textit{``LLM Score''}). We then combined the two scores by averaging them, creating a single score for each metric as shown in Equation \ref{eqn:even-eval}. The \textit{Metric Score} refers to the score of any one of the four metrics from Section \ref{sec:hm_metrics}.

\begin{equation}
\label{eqn:even-eval}
\resizebox{0.5\hsize}{!}{$Metric \: Score = Sim \: Score\times 4 \times 0.5 + LLM \: Score \times 0.5$}
\end{equation}

%To do so, we first normalized the evaluation score of \textit{Retrieval-based Quality Evaluator} (denoted by \textit{``Sim Score''}. It ranges from 0 to 1, however to compare with \textit{LLM-based Quality Evaluator} (denoted by \textit{``LLM Score''}), it needs to be changed to a scale of 0 to 4. We achieve this by multiplying \textit{``Sim Score''} by 4. Then, we evenly weighted the two scores into one score for each metric as shown in Equation \ref{eqn:even-eval} (\textit{Metric Score} represents the score of any one of the four metrics in Section \ref{sec:hm_metrics}). 

Additionally, we propose that the combined evaluation score may better approximate human judgment if we factor in the correlation between \textit{``Sim Score’’} and human-labeled scores, and \textit{``LLM Score’’} and human-labeled scores in the weighting strategy. Similar to prior research using Pearson correlation to assess the effectiveness of automated metrics compared to human judgment \cite{wang2023chatgpt,liu2023g}, we performed a Pearson correlation analysis on the scores from the \textit{LLM-based Quality Evaluator} in the testing split. The correlation coefficients were then used to weight \textit{``Sim Score’’} and \textit{``LLM Score’’} in the combined score (Equation \ref{eqn:corl-eval}). \textit{Sim Coeff} represents the correlation coefficient between \textit{``Sim Score’’} and human-labeled scores, while \textit{``LLM Coeff’’} represents the same for \textit{``LLM Score’’}. For the \textit{Retrieval-based Quality Evaluator}, we set it to retrieve the top 10 most similar git diffs, as this configuration produced the highest correlation coefficient. Notably, since \textit{``Sim Score’’} did not show a significant correlation (p-value > 0.05) with human-labeled scores for \textit{Conciseness}, we relied solely on \textit{``LLM Score’’} for that metric.

%In addition, we posit that the combined evaluation score can more effectively approximate the human judgement if we consider the correlation between \textit{``Sim Score''} and human-labeled scores, and \textit{``LLM Score''} and human-labeled scores into the weighting strategy. Thus, similar to previous works that used Pearson correlation to assess the effectiveness of automated metrics compared to human judgement \cite{wang2023chatgpt,liu2023g}, we first conducted Pearson correlation \cite{sedgwick2012pearson} on the scores that evaluated messages in the testing split of \textit{LLM-based Quality Evaluator}, and used the correlation coefficients to weight \textit{``Sim Score''} and \textit{``LLM Score''} in the combined score as shown in Equation \ref{eqn:corl-eval}. \textit{Sim Coeff} represents the correlation coefficient value between \textit{``Sim Score''} and human-labeled scores. Similarly, \textit{LLM Coeff} represents the correlation coefficient value between \textit{``LLM Score''} and human-labeled scores. Point to note, for \textit{Retrieval-based Quality Evaluator}, we set it to retrieve top 10 most similar git diffs because this setting showed the highest correlation coefficient among all other settings. Since \textit{``Sim Score''} under any settings did not show correlation (p-value > 0.05) with human-labeled scores for \textit{Conciseness}, we only relied on \textit{``LLM Score''} for \textit{Conciseness}.

\begin{equation}
\label{eqn:corl-eval}
\resizebox{0.75\hsize}{!}{$Metric \: Score = Sim \: Score\times 4 \times (\frac{Sim \: Coeff}{Sim \: Coeff + LLM \: Coeff}) + LLM \: Score \times (\frac{LLM \: Coeff}{Sim \: Coeff + LLM \: Coeff}) $}
\end{equation}

%It ranges from 0 to 1 as we found positive correlations, least to most semantically similar between generated and human-written messages) to the scale of 0 to 4, while keeping the evaluation score of \textit{LLM-based Quality Evaluator} (denoted by \textit{``LLM Score''}) at the same scale. 

%\textit{Sim Coeff} represents the coefficient value of \textit{``Sim Score''} and \textit{LLM Coeff} is for \textit{``LLM Score''}). 

%\textit{Sim Coeff} represents the correlation coefficient value between \textit{``Sim Score''} and human-labeled scores. Similarly, \textit{LLM Coeff} represents the correlation coefficient value between \textit{``LLM Score''} and human-labeled scores.

\subsection{Search-based Optimization Algorithm Design}
\label{sec:opt}
Algorithm~\ref{alg:cap} outlines the search-based optimization process to enhance the quality of human-written commit messages. The initial variables are defined in lines 1–4, including the current search step number (\textit{step}) and the maximum number of allowed steps (\textit{step\_limit}). We used the tools described in Section \ref{sec:tools} to retrieve relevant software contexts. Additionally, we incorporated all tools from OMG, such as the \textit{pull request/issue report title retriever} and the \textit{commit type classifier}. A comprehensive list of all automatically retrieved contexts is provided in Table \ref{tab:contexts}.

%OLD Algorithm~\ref{alg:cap} illustrates the search-based optimization for maximizing the quality of human-written commit messages. Initial variables are specified in line 1-4, including the current search step number (\textit{step}), the maximum number of steps allowed in the optimization (\textit{step\_limit}). To retrieve available contexts, we used the tools in Section \hl{missing} \ref{sec:tools}. In addition, we also incorporated all the information retrieval tools from OMG such as \textit{pull request/issue report title retriever} and \textit{commit type classifier}. All the software contexts that we automatically retrieved by tools are presented in Table \ref{tab:contexts}. 

Since the optimization begins with a human-written commit message, we improved the performance of the \textit{commit type classifier} from OMG by including both the git diff and the human-written message in the classifier prompt. Following OMG, we evaluated the classifier on the dataset provided by Levin et al. \cite{levin2017boosting}, using accuracy as the evaluation metric, as both Levin et al. and OMG reported only accuracy. With the inclusion of human-written messages, we achieved an accuracy of 81.0\%, surpassing OMG's classifier (53.6\%) and Levin et al.'s results (48.8\%-54.3\% without commit messages, 70.3\%-76.7\% with commit messages). Given that commit type information is a key part of the format expected by developers \cite{li2024only}, we provided this information directly rather than using it as a context that could be used at each optimization step. Thus, we finally have seven software contexts in total that could be considered to optimize messages.

%OLD: Since the optimization starts from a human-written message, we enhanced the performance of \textit{commit type classifier} from OMG by including both git diff and its human-written message in the prompt to the classifier. Following OMG, we evaluated the classifier's performance on a dataset provided by Levin et al. \cite{levin2017boosting}. Since both Levin et al. and OMG only reported the accuracy of their classifiers, we also utilized accuracy in this study. With the help of human-written messages, we achieved an accuracy of 81.0\%, which outperformed both \textit{commit type classifier} in OMG (53.6\%) and by Levin et al. (48.8\%-54.3\% without commit message as input, 70.3\%-76.7\% with commit message). Finally, due to the fact that commit type information is an indispensable part of the commit message writing format expected by developers \cite{li2024only}, we directly provided such information instead of treating it as a context that could be used at each step of the optimization. \hl{Thus, we have seven software contexts in total that could be considered to optimize messages.}

%Moreover, we retrieved the discussion comments of pull request/issue report in addition to their titles for more detailed context information. 

\begin{table}[]
\footnotesize
\setlength{\tabcolsep}{2pt}
\caption{Automatically Retrieved Software Contexts}
\label{tab:contexts}
\begin{tabular}{c|c}
\hline
Important File Information & Commit Type Information      \\ \hline
Pull Request/Issue Report  & Method Body Summary           \\ \hline
Class Body Summary        & Syntactically Related Blocks  \\ \hline
Invoked Methods  & Variable Data Types              \\ \hline
\end{tabular}
\end{table}

% \begin{table}[]
% \footnotesize
% \setlength{\tabcolsep}{2pt}
% \caption{Automatically Retrieved Software Contexts}
% \label{tab:contexts}
% \begin{tabular}{c|c}
% \hline
% Important File Information & Commit Type Information      \\ \hline
% Pull Request/Issue Report  & Pull Request/Issue Title           \\ \hline
% Method Body Summary        & Class Body Summary  \\ \hline
% Syntactically Related Blocks  & Invoked Methods              \\ \hline
% Variable Data Types        &  Discussion comments of Pull Request/Issue Report\\ \hline
% \end{tabular}
% \end{table}

To simplify our search-based algorithm for optimizing a single objective function, we summed the four metrics from the \textit{Combined Evaluation Score} (Section \ref{sec:obj}, referred to as the \textit{optimization score}, which is returned by the \textit{EVALUATE} function.) The \textit{priority queue} is sorted by this optimization score, ensuring that the highest-quality candidate remains at the front (line 16). The variable \textit{highest\_score} tracks the top optimization score so far, initialized with the score of the human-written message. The algorithm first updates the human-written commit message using each available context individually (Table \ref{tab:contexts}), generating different commit message candidates (\textit{msg\_candidate}). At each subsequent step, the candidate with the highest optimization score is dequeued from the \textit{priority queue} and updated with the contexts that haven't been considered yet, generating further candidates for improvement (lines 11-17).

%OLD To simplify our search-based algorithm to optimize a single value as the objective function, we summed the scores of the four metrics from \textit{Combined Evaluation Score} in Section \ref{sec:obj} (denoted as \textit{optimization score}, returned by \textit{EVALUATE} function), and sorted the \textit{priority queue} by this optimization score to ensure the candidate with the highest quality stays at the front (line 16). \textit{highest\_score} keeps track of the highest optimization score so far, which is initialized with the optimization score of human-written message. The algorithm first updates the human-written commit message with all available contexts individually (Table \ref{tab:contexts}) to generate different commit message candidates (\textit{msg\_candidate}). At each following step, the candidate with the highest optimization score is dequeued from the \textit{priority queue} and updated with the contexts individually that have not been considered to generate candidates for further updating (line 11-17). 

We implemented \textit{UPDATE} function through prompting GPT-4 as it has shown state-of-the-art performance in CMG by prompting \cite{li2024only}. In the prompt, we included the target git diff, the definition of a git diff, the expected commit message format, and explanations of the four metrics and their scoring criteria \cite{li2024only}. To help GPT-4 optimize the scores from the \textit{Retrieval-based Quality Evaluator} (part of the overall optimization score), we also provided the top 10 git diffs most similar to the target diff, along with their corresponding commit messages. GPT-4 was instructed to improve the existing commit message (whether human-written or a previous candidate) rather than generating an entirely new message.

Specifically, each new candidate commit message is refined by considering one additional software context from Table \ref{tab:contexts} that had not yet been addressed. To facilitate this, we prompted GPT-4 with the new context, the previous candidate message, the contexts already considered, the evaluation scores from the \textit{Combined Evaluation Score}, and instructed the model to consider the new context to generate another candidate that is of higher quality than the previous candidate in terms of the evaluation scores.

%To ensure GPT-4 can optimize the scores by \textit{Retrieval-based Quality Evaluator} (as part of the optimization score), we also provided the top 10 git diffs that are the most similar to the target diff along with their commit messages in the prompt. Then, we explicitly instructed it to improve upon the existing message (either human-written or a previous candidate message) instead of completely abandoning it to generate a new candidate. 

%Specifically, a new candidate tries to improve upon the previous one by considering one additional software context from Table \ref{tab:contexts} that it has not considered before. To do so, we prompted GPT-4 with the new context, the previous candidate, the contexts that have been considered by this candidate in the past iterations, its evaluation scores from \textit{Combined Evaluation Score}, and instructed the model to consider the new context to generate another candidate that is of higher quality than the previous candidate in terms of the evaluation scores. 

We opted to introduce only one context at each step because a software context may be highly relevant to certain code changes but marginal or irrelevant to others, depending on the specific nature and motivation behind the changes. Studies show that LLM performance declines when irrelevant context is included \cite{jones2022capturing,shi2023large}, as it becomes harder for the model to discern the most important parts of the prompt. Additionally, OMG relied entirely on GPT-4 to reason about the retrieved information and select relevant contexts, but research shows that LLMs can make reasoning errors and incorrect judgments \cite{stechly2023gpt,huang2023large}. By considering one context at a time, we maintain finer control over the optimization process with feedback from our evaluators.

%The reason for considering only one context at each step is: \textit{we posit that a software context may be heavily relevant to some code changes but marginally relevant (or even irrelevant) to others, depending on the exact changes in the code and the motivation behind the changes}. Since the performance of LLMs decreases when irrelevant context is included \cite{jones2022capturing,shi2023large} in the prompt, incorporating all available contexts (if they can be readily retrieved) may negatively impact the quality of the generated messages by making the LLMs hard to discern which part of the prompt to focus on. Moreover, OMG completely relied on GPT-4 to reason about the retrieved information and decide which context to retrieve next or generate the final message, while numerous studies have shown that LLMs could make reasoning mistakes and incorrect judgement \cite{stechly2023gpt,huang2023large}. Thus, considering one context at a time would grant a fine-grained control of the optimization trajectory with the feedback from our evaluators. 

To stop the optimization process, we incorporated multiple stopping criteria alongside the \textit{step\_limit}. First, we introduced a dynamic score improvement threshold (\textit{improve\_threshold}) that decreases as the number of iterations grows (line 7). This approach assumes that the degree of improvement should naturally diminish over time as the candidate messages improve in quality. Initially, the threshold is set to be a percentage (\textit{p}) of the optimization score of the human-written message (line 3), with a minimum threshold (\textit{min\_threshold}) in place to prevent it from approaching zero (line 8-9). The optimization halts if the score improvement between the newly updated \textit{highest\_score} and the \textit{highest\_score} updated two times ago (i.e., we keep track of the \textit{highest\_score} every time it's updated) is less than the threshold (line 22-23). Additionally, we set GPT-4’s temperature to zero to generate deterministic outputs, following previous research \cite{peng2023towards,li2024only}. However, when the improvement between the newly updated \textit{highest\_score} and the \textit{highest\_score} updated two times ago is less than the threshold, we raise the temperature. This allows GPT-4 to generate a new candidate with greater variability, potentially exceeding the threshold and continuing the optimization process \cite{liu2023large}.

%To stop the optimization, we designed several stopping criteria in addition to \textit{step\_limit}. First, a score improvement threshold (\textit{improve\_threshold}) was deployed that decreases as the number of iterations increases (line 7), since we posit that the improvement should naturally become smaller over time as the candidates have better quality. In line 3, we initialize the threshold to be \textit{p} percent of the human-written message's optimization score, and the minimum threshold, \textit{min\_threshold}, is also set to prevent it from being close to zero (line 8-9). The optimization is stopped once the increase of the optimization score in two consecutive steps is not larger than the threshold (line 22-23). In addition, we set the temperature of GPT-4 to be zero for the most deterministic generations following previous works \cite{peng2023towards,li2024only}. However, if the increase of the optimization score in two consecutive steps is not larger than the threshold, we increased the temperature to let GPT-4 generate the candidate again as the LLMs generate more non-deterministic answers with higher temperatures, which may be a better candidate whose score is larger than the threshold so that the optimization continues \cite{liu2023large}. 

\begin{algorithm}[hbt!]
\caption{Search-based Optimization}\label{alg:cap}
\begin{algorithmic}[1]
\Require{git diff, available contexts, human msg}
\Ensure{optimized msg, evaluation score}
\State $step \gets 0$, $step\_limit \gets N$
\State $highest\_score \gets $ \texttt{EVALUATE(git diff, human msg)}
\State $improve\_threshold \gets highest\_score \times p$, $min\_threshold \gets \frac{improve\_threshold}{step\_limit}$
\State $priority\_queue$\texttt{.enqueue(human msg)}
\While{$step \textless step\_limit$}
    \State $step \gets step+1$
    \State $improve\_threshold \gets improve\_threshold - \frac{(improve\_threshold \times step)}{step\_limit}$
    \If{$improve\_threshold \textless min\_threshold$}
        \State $improve\_threshold \gets min\_threshold$
    \EndIf
    \State $cur\_msg \gets priority\_queue${.dequeue()}
    \For{\texttt{each context in available contexts}}
        \State $msg\_candidate \gets $ \texttt{UPDATE($cur\_msg$, git diff, context, considered contexts, EVALUATE(git diff, $cur\_msg$))}
        \State $priority\_queue$\texttt{.enqueue($msg\_candidate$)}
    \EndFor
    \State $priority\_queue$\texttt{.sort(key = evaluation score)}
    \State $cur\_score \gets $ \texttt{EVALUATE(git diff, $priority\_queue$[0].msg)}
    \If{$cur\_score \textgreater highest\_score$}
        \State $highest\_score \gets cur\_score$
        \State optimized msg $ \gets $ \texttt{$priority\_queue$[0].msg}
        \State evaluation score $ \gets highest\_score $
        \If{$highest\_score - highest\_score_{updated\_two\_iterations\_ago}$ $\textless improve\_threshold$}
            \State \texttt{break}
        \EndIf
    \EndIf
\EndWhile
\end{algorithmic}
\end{algorithm}

\subsection{Hyper-parameter Tuning}
\label{sec:hypm}

As shown in Algorithm \ref{alg:cap}, several hyper-parameters that may alter the behavior of CMO exist, such as \textit{step\_limit} and percentage value \textit{p}. Also, the value of the increased temperature (Section \ref{sec:opt}) can be set to control GPT-4's randomness in the optimized messages when the improvement in scores is not larger than the threshold. In addition, the selection of the equations (Equation \ref{eqn:even-eval} or \ref{eqn:corl-eval}) for calculating \textit{Combined Evaluation Score} (Section \ref{sec:obj}) may affect the performance. 

To optimize CMO's performance, we tested a range of hyper-parameters: \textit{p} (5, 10, 15, 20), \textit{temperature} (0.5, 1), \textit{step\_limit} (10, 30, 50, 60), and two equations for calculating the \textit{Combined Evaluation Score}. Using a Grid Search approach\cite{jimenez2008finding}, we evaluated all 64 combinations of these hyper-parameters. Due to budget constraints with the OpenAI API \cite{openaiapi}, we randomly sampled 10 commits and optimized their human-written messages. Four authors independently evaluated the results using the metrics in Section \ref{sec:hm_metrics}. Based on these evaluations, we selected \textit{p} = 5, temperature = 1, \textit{step\_limit} = 50, and Equation \ref{eqn:corl-eval} for the final CMO settings used in the following sections.

%OLD:To select the hyper-parameters that maximize CMO's performance, we experimented with a range of values for \textit{p} (5, 10, 15, and 20), increased temperature (0.5 and 1), \textit{step\_limit} (10, 30, 50, 60), and the two equations for calculating \textit{Combined Evaluation Score}. Similar to Grid Search \cite{jimenez2008finding} in Machine Learning, we optimized the human-written messages using all combinations of these hyper-parameters (64 in total). Due to our limited financial budget of calling OpenAI API, we randomly sampled ten commits from the testing split of \textit{LLM-based Quality Evaluator} (Section \ref{sec:obj}) and optimized their human-written messages. Then, four of the authors manually and independently examined the messages by considering the four metrics in Section \ref{sec:hm_metrics}. Finally, based on our judgement of the optimized message quality, we set \textit{p} as 5, increased temperature as 1, \textit{step\_limit} as 50, and used Equation \ref{eqn:corl-eval}. This setting of CMO is adopted in the following sections. 

%\hl{Jiawei: can we simply use magic numbers and put using grid search in TTV? Is this good enough? reviewers might find issues in this section. IA: looks good to me. TTV updated}

\section{RQ2: The Effectiveness of CMO}
\label{sec:rq2}
This section presents the performance of CMO compared to the baselines (Section \ref{sec:baselines}). Table \ref{tab:researcher-evalrank} shows the results of the researchers' ranking of CMO and the three baseline messages. Each cell indicates how many times messages from a specific baseline were ranked in a certain position for a given metric. With two researchers ranking four messages across 76 commits, each column sums to 152. The highest numbers in each column are highlighted: CMO-optimized messages ranked first 135 times (88.8\%) for \textit{Rationality}, 134 times (88.2\%) for \textit{Comprehensiveness}, and 145 times (95.4\%) for \textit{Expressiveness}, outperforming OMG, CMC, and human-written messages. However, CMO was ranked first only 63 times (41.4\%) in terms of \textit{Conciseness}.
%OLD: In this section, we present the performance of CMO compared with the baselines (Section \ref{sec:baselines}). We first present the results of the researchers' ranking of the messages by CMO and the three baselines in Table \ref{tab:researcher-evalrank}. Each cell in the table represents the number of times the messages from a certain baseline (human is denoted as HM) was ranked at a specific position based on a metric. The sum of each column is 152 as two researchers were asked to rank the four messages associated with 76 commits from the testing split of \textit{LLM-based Quality Evaluator} (Section \ref{sec:obj}). The largest number in each column is highlighted: CMO-optimized messages were ranked the best (i.e., first position) 135 times (88.8\%) in terms of \textit{Rationality}, 134 times (88.2\%) in \textit{Comprehensiveness}, and 145 times (95.4\%) in \textit{Expressiveness}, which are larger than that of OMG, CMC, and human. This shows that CMO effectively optimizes the human-written messages and outperforms both OMG and CMC from the researchers' perspective. In addition, both of them ranked CMO-optimized messages at top for 80.3\% of the commits in \textit{Rationality}, 76.4\% in \textit{Comprehensiveness}, and 90.9\% in \textit{Expressiveness}. However, CMO-optimized messages falls behind in terms of \textit{Conciseness} with only 22.2\% of the commits were ranked at the top by both researchers. 

Table \ref{tab:developer-eval} presents the evaluation results from Apache developers. Each cell shows how often messages from a specific baseline were ranked in a particular position, with each column totaling 220 since 22 developers ranked four messages across 10 commits. The results indicate that CMO-optimized messages ranked first 143 times (65.0\%) for \textit{Rationality}, 139 times (63.2\%) for \textit{Comprehensiveness}, and 85 times (38.6\%) for \textit{Expressiveness}, aligning with the researchers' evaluation. This reinforces the finding that CMO outperforms OMG/CMC and enhances human-written messages, though it still lags behind in \textit{Conciseness}.

%OLD:Table \ref{tab:developer-eval} shows the evaluation results from Apache developers. Similarly, each cell in the table represents the number of times the messages from a certain baseline was ranked at a specific position. The sum of each column is 220 as 22 developers were asked to rank the four messages associated with 10 commits. From the results, CMO-optimized messages were ranked the best 143 times (65.0\%) in terms of \textit{Rationality}, 139 times (63.2\%) in \textit{Comprehensiveness}, and 85 times (38.6\%) in \textit{Expressiveness}, which shows a pattern similar to researchers' evaluation. This further supports the results from the researchers that CMO outperforms OMG/CMC and optimizes the quality of human-written messages. However, we notice that its \textit{Conciseness} is worse than the three baselines. 

\begin{table}[]
\footnotesize
\setlength{\tabcolsep}{1pt}
\caption{Human Evaluation (by researchers) Results on Commit Messages}
\label{tab:researcher-evalrank}
\begin{tabular}{l|cccc|cccc|cccc|cccc}
\hline
\multicolumn{1}{c|}{} & \multicolumn{4}{c|}{\textbf{Rationality}}                & \multicolumn{4}{c|}{\textbf{Comprehensiveness}}          & \multicolumn{4}{c|}{\textbf{Conciseness}}                & \multicolumn{4}{c}{\textbf{Expressiveness}}              \\ \hline
\multicolumn{1}{c|}{} & \textbf{CMO} & \textbf{OMG} & \textbf{CMC} & \textbf{HM} & \textbf{CMO} & \textbf{OMG} & \textbf{CMC} & \textbf{HM} & \textbf{CMO} & \textbf{OMG} & \textbf{CMC} & \textbf{HM} & \textbf{CMO} & \textbf{OMG} & \textbf{CMC} & \textbf{HM} \\ \hline
\textbf{Rank 1st}     & \textbf{135} & 28           & 16           & 21          & \textbf{134} & 44           & 16           & 17          & 63           & 52           & 71           & \textbf{74} & \textbf{145} & 95           & 64           & 52          \\
\textbf{Rank 2nd}     & 12           & \textbf{88}  & 9            & 30          & 15           & \textbf{85}  & 9            & 15          & 14           & \textbf{57}  & 25           & 17          & 6            & \textbf{55}  & 2            & 5           \\
\textbf{Rank 3rd}     & 3            & 22           & \textbf{84}  & 37          & 1            & 16           & \textbf{90}  & 40          & 21           & \textbf{37}  & 35  & 27          & 0            & 2            & \textbf{52}  & 29          \\
\textbf{Rank 4th}     & 2            & 14           & 43           & \textbf{64} & 2            & 7            & 37           & \textbf{80} & \textbf{54}  & 6            & 21           & 34          & 1            & 0            & 34           & \textbf{66} \\ \hline
\end{tabular}
\end{table}

\begin{table}[]
\footnotesize
\setlength{\tabcolsep}{1pt}
\caption{Human Evaluation (by developers) Results on Commit Messages}
\label{tab:developer-eval}
\begin{tabular}{l|cccc|cccc|cccc|cccc}
\hline
\multicolumn{1}{c|}{} & \multicolumn{4}{c|}{\textbf{Rationality}}                 & \multicolumn{4}{c|}{\textbf{Comprehensiveness}}           & \multicolumn{4}{c|}{\textbf{Conciseness}}                 & \multicolumn{4}{c}{\textbf{Expressiveness}}              \\ \hline
\multicolumn{1}{c|}{} & \textbf{CMO} & \textbf{OMG} & \textbf{CMC} & \textbf{HM}  & \textbf{CMO} & \textbf{OMG} & \textbf{CMC} & \textbf{HM}  & \textbf{CMO} & \textbf{OMG} & \textbf{CMC} & \textbf{HM}  & \textbf{CMO} & \textbf{OMG} & \textbf{CMC} & \textbf{HM} \\ \hline
\textbf{Rank 1st}     & \textbf{143} & 43           & 12           & 19           & \textbf{139} & 56           & 11           & 11           & 15           & 26           & 82           & \textbf{100} & \textbf{85}  & 71           & 35           & 27          \\
\textbf{Rank 2nd}     & 43           & \textbf{130} & 29           & 20           & 48           & \textbf{131} & 29           & 12           & 19           & 25           & \textbf{100} & 77           & 60           & \textbf{92}  & 29           & 36          \\
\textbf{Rank 3rd}     & 19           & 25           & \textbf{100} & 76           & 17           & 19           & \textbf{124} & 60           & 54           & \textbf{135} & 16           & 15           & 41           & 32           & \textbf{90}  & 58          \\
\textbf{Rank 4th}     & 15           & 22           & 79           & \textbf{105} & 16           & 14           & 56           & \textbf{137} & \textbf{132} & 34           & 22           & 28           & 34           & 25           & 66           & \textbf{99} \\ \hline
\end{tabular}
\end{table}

We also evaluated the performance using automated evaluators by calculating the average scores for the \textit{Combined Evaluation Score} with Equation \ref{eqn:corl-eval} (Section Section \ref{sec:obj}), averaged across all commits in the testing split of the \textit{LLM-based Quality Evaluator}. We conducted Welch's t-test \cite{ruxton2006unequal} and Cohen's D \cite{diener2010cohen} to assess the statistical significance of score differences between approaches. As shown in Table \ref{tab:evaluators-eval}, CMO statistically significantly outperforms OMG (\textit{Rationality}: p-val<2.961e-7, Cohen’s D (0.887, large); \textit{Comprehensiveness}: p-val<2.750e-6, Cohen’s D (0.805, large); \textit{Expressiveness}: p-val<2.961e-7, Cohen’s D (0.887, large)) and CMC (\textit{Rationality}: p-val<6.423e-6, Cohen’s D (0.775, large); \textit{Comprehensiveness}: p-val<1.773e-13, Cohen’s D (1.346, large); \textit{Expressiveness}: p-val<7.965e-13, Cohen’s D (1.302, large)), indicating CMO produces higher-quality messages than state-of-the-art generation/completion techniques. Additionally, CMO effectively optimizes human-written messages (\textit{Rationality}: p-val<1.698e-9, Cohen’s D(1.065, large); \textit{Comprehensiveness}: p-val<4.478e-6, Cohen’s D (0.789, large); \textit{Expressiveness}: p-val<6.032e-13, Cohen’s D (1.310, large)).

Aligning with the findings of Li et al. \cite{li2024only}, our analysis using traditional automatic evaluation metrics (Section \ref{sec:bleu}) does not reflect the performance difference between CMO and CMC. CMC scored significantly higher in BLEU, METEOR, and ROUGE-L (e.g., BLEU for CMC is 21.40, while for CMO, it's only 2.91). These metrics rely on human-written messages as references, so they cannot gauge whether the CMO has effectively optimized these messages. This underscores the limitations of traditional metrics in evaluating CMG-related tasks, corroborating with the findings of Li et al. \cite{li2024only}.

%OLD However, the traditional automatic evaluation metrics (Section \ref{sec:bleu}) do not reflect such performance difference as CMC achieved significantly higher scores than CMO in BLEU, METEOR, and ROUGE-L (i.e., BLEU for CMC is 21.40 while it's 2.91 for CMO), and we can not gauge whether CMO has optimized the human-written messages as they are used as the references by these metrics. This indicates the ineffectiveness of traditional metrics in evaluating CMG-related tasks similar to the findings by Li et al. \cite{li2024only}. 

\begin{mdframed}[roundcorner=10pt]
\textbf{Finding 1: CMO effectively optimizes the human-written commit messages and outperforms state-of-the-art CMG/CMC techniques in \textit{Rationality}, \textit{Comprehensiveness}, and \textit{Expressiveness}.} 
\end{mdframed}

While CMO demonstrates superior performance over OMG/CMC and effectively optimizes human-written messages, its \textit{Conciseness} remains lacking. Prior research indicates that LLMs tend to generate more detailed and longer commit messages compared to humans \cite{eliseeva2023commit,zhang2024automatic, li2024only}, which may explain this outcome.

%OLD: This may be due to the potential bias in the \textit{LLM-base Quality Evaluator} as it was solely relied on to evaluate \textit{Conciseness} (Section \ref{sec:obj}). A larger and more comprehensive manually labeled training dataset may be needed to train LLMs to be more effective in evaluating \textit{Conciseness}. We leave constructing such a benchmark as a future work. Moreover, since researchers have shown that LLMs tend to generate commit messages that are more detailed and longer than human-written messages \cite{eliseeva2023commit,zhang2024automatic, li2024only}, a new evaluation metric should be proposed for better evaluating the \textit{Conciseness} of LLM-generated commit messages.

\begin{figure*}[t]
     \centering
     \includegraphics[width=\textwidth]{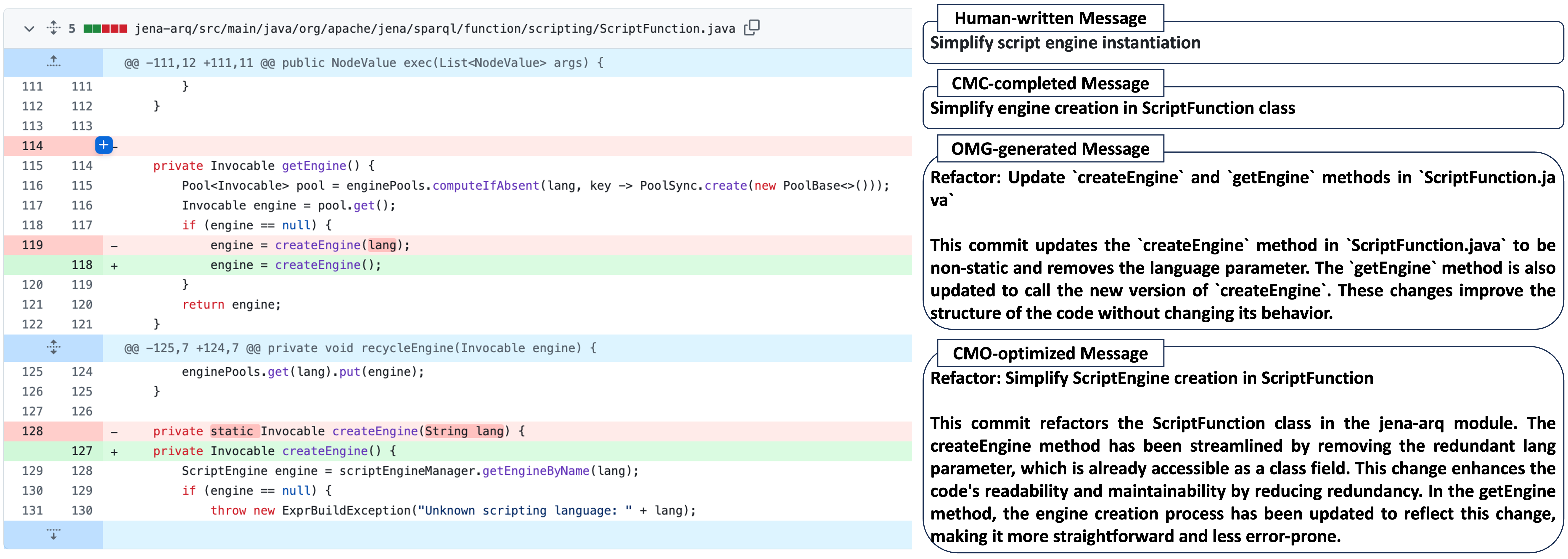}
     \caption{Example of Code Refactoring}
     \label{fig:success}
\end{figure*}

Figure \ref{fig:success} \cite{successfulcase} illustrates an example where CMO optimizes a human-written commit message and outperforms both OMG and CMC, as confirmed by researchers and automated evaluators. In this case, the human-written and CMC-completed messages merely note that the code changes simplify the method definition of \textit{createEngine}, without explaining why the \textit{lang} parameter can be removed without affecting functionality. OMG provides a detailed description of the ``What'' aspect (e.g., \textit{``...to be non-static and removes the language parameter...''}, \textit{``...also updated to call the new version of createEngine...''}) but fails to explain why the parameter is removed, only mentioning general refactoring (\textit{``...improve the structure of the code...''}). The CMO-optimized message, not only provides a clear description of the ``What'' (\textit{``...streamlined by removing the redundant lang parameter...''}), but also explains the rationale (\textit{.``..enhances the code's readability and maintainability by reducing redundancy...''}) and the reason for the change (\textit{``...which is already accessible as a class field...''}).

\begin{table}[]
\footnotesize
\setlength{\tabcolsep}{1pt}
\caption{Automatic Evaluation Results (by evaluators) on Commit Messages}
\label{tab:evaluators-eval}
\begin{tabular}{l|ccccc}
\hline
                    & \textbf{Rationality} & \textbf{Comprehensiveness} & \textbf{Conciseness} & \textbf{Expressiveness} & \textbf{Total}  \\ \hline
\textbf{CMO}        & \textbf{3.054}       & \textbf{3.283}             & 4                    & \textbf{3.274}          & \textbf{13.611} \\
\textbf{OMG}        & 2.942                & 3.178                      & 4                    & 3.188                   & 13.308          \\
\textbf{CMC}        & 2.602                & 3.141                      & 4                    & 3.142                   & 12.884          \\
\textbf{Human}         & 2.097                & 2.692                      & 4                    & 3.129                   & 11.918          \\
\hline
\textbf{CMO-Search} & 2.965                & 3.215                      & 4                    & 3.206                   & 13.386          \\
\textbf{CMO-File}   & 2.937                & 3.198                      & 4                    & 3.240                   & 13.375          \\
\textbf{CMO-Type}   & 3.033                & 3.266                      & 4                    & 3.258                   & 13.557          \\
\textbf{CMO-MB}     & 3.050                & 3.279                      & 4                    & 3.271                   & 13.600          \\ \hline
\end{tabular}
\end{table}

\section{RQ3: Ablation Study}
\label{sec:rq3}
We conducted an ablation study to assess the impact of the CMO's key components: the automated software context collection tools and the search-based optimization, which includes the automated evaluators (i.e., the objective function). Due to financial constraints in calling the OpenAI API \cite{openaiapi}, we couldn't remove all seven context collection tools (Section \ref{sec:opt}) individually to create seven distinct CMO variants. Instead, we identified the three most frequently used tools in Section \ref{sec:rq2}: \textit{Important File Information}, \textit{Method Body Summary}, and \textit{Variable Data Types}. We created three CMO variants–\textit{CMO-File}, \textit{CMO-MethodBody}, and \textit{CMO-Type}--where each variant excluded one of these tools but retained access to the other six.

For search-based optimization, we removed the search algorithm and included all contexts retrieved by the seven tools directly in the prompt for GPT-4 to optimize human-written messages. This variant was labeled \textit{CMO-Search}.

%We conducted an ablation study to investigate the effectiveness of each major component in CMO, namely, the automated software context collection tools and the search-based optimization that includes the automated evaluators (i.e., objective function). 

%For software context collection tools, due to our financial restrictions of calling OpenAI API, we could not remove the seven tools (Section \ref{sec:opt}) one at a time to create seven different CMO variants. Instead, we first identified the three tools most frequently used by CMO in Section \ref{sec:rq2}. As a result, these tools are the ones that retrieve \textit{Important File Information}, \textit{Method Body Summary}, and \textit{Variable Data Types}. Then, we created three CMO variants where each of them has access to all the seven tools except the three frequently used ones. We named them as \textit{CMO-File}, \textit{CMO-MethodBody}, and \textit{CMO-Type}, depending on which tool was eliminated. 

%For search-based optimization, we simply removed the search algorithm and put all the contexts returned by the seven tools in the prompt for GPT-4 to optimize the human-written messages. We denoted this variant as \textit{CMO-Search}.

As shown in Table \ref{tab:evaluators-eval}, removing one of the frequently used tools reduces CMO's average scores in \textit{Rationality}, \textit{Comprehensiveness}, and \textit{Expressiveness} compared to CMO. However, these differences are not statistically significant (p-value > 0.05) when evaluated against \textit{CMO-File}, \textit{CMO-MethodBody}, and \textit{CMO-Type}. In contrast, \textit{CMO-Search} shows statistically significant lower average scores than CMO in all metrics: \textit{Rationality} (Welch’s t-test, p-value < 9.034e-5, Cohen’s D = 0.667, medium), \textit{Comprehensiveness} (Welch’s t-test, p-value < 9.034e-5, Cohen’s D = 0.667, medium), and \textit{Expressiveness} (Welch’s t-test, p-value < 9.034e-5, Cohen’s D = 0.667, medium).

Table \ref{tab:researcher-eval} presents the researchers' evaluations, highlighting the relative rankings where both researchers agreed. The table indicates whether CMO-optimized messages were ranked \textit{Higher}, \textit{Equal}, or \textit{Lower} compared to other variants. The highest percentages for each metric are highlighted.
For comparisons between CMO and \textit{CMO-File}, \textit{CMO-MethodBody}, and \textit{CMO-Type}, CMO-optimized messages were considered better by both researchers for 14.1\%-23.9\% of commits in terms of \textit{Rationality}, 4.2\%-11.1\% in terms of \textit{Comprehensiveness}, 6.9\%-11.1\% in terms of \textit{Conciseness}, and 1.3\%-2.7\% in terms of \textit{Expressiveness}. This indicates that removing different context collection tools affects \textit{Rationality} more significantly than the other metrics. Additionally, both researchers found that \textit{CMO-Search} negatively impacted more commits (25.4\% in \textit{Rationality}) across all metrics compared to the removal of individual context collection tools.

The results suggest that selectively choosing which contexts to include is more effective for optimizing message quality compared to including all available contexts in the prompt (\textit{CMO-Search}). Removing access to a single context from CMO (\textit{CMO-File}, \textit{CMO-MethodBody}, or \textit{CMO-Type}) has a smaller overall impact on performance since missing one context affects the quality of generated messages for some commits but not others. For example, removing \textit{Method Body Summary} would not impact messages for commits modifying code outside of any methods.

\begin{mdframed}[roundcorner=10pt]
\textbf{Finding 2: Different context collection tools impact the quality of the generated commit messages in varying ways, with \textit{Rationality} being most sensitive to changes.} 
\end{mdframed}

\begin{table}[]
\footnotesize
\setlength{\tabcolsep}{0.8pt}
\caption{Human Evaluation (by researchers) Results on Commit Messages}
\label{tab:researcher-eval}
\begin{tabular}{l|cccccccccccc}
\hline
                           & \multicolumn{3}{c}{\textbf{Rationality}}          & \multicolumn{3}{c}{\textbf{Comprehensiveness}}     & \multicolumn{3}{c}{\textbf{Conciseness}}           & \multicolumn{3}{c}{\textbf{Expressiveness}}        \\ \cline{2-13} 
                           & \textbf{Higher} & \textbf{Equal} & \textbf{Lower} & \textbf{Higher} & \textbf{Equal}  & \textbf{Lower} & \textbf{Higher} & \textbf{Equal}  & \textbf{Lower} & \textbf{Higher} & \textbf{Equal}  & \textbf{Lower} \\ \hline
\textbf{CMO VS CMO-Search} & \textbf{25.4\%} & 2.8\%          & 4.2\%          & 11.1\%          & 11.1\%          & 6.9\%          & \textbf{15.3\%} & 12.5\%          & 4.2\%          & 4.0\%           & \textbf{5.3\%}  & 0.0\%          \\
\hline
\textbf{CMO VS CMO-File}   & \textbf{23.9\%} & 2.8\%          & 11.3\%         & 11.1\%          & 11.1\%          & 8.3\%          & 6.9\%           & 8.3\%           & 8.3\%          & 1.3\%           & \textbf{4.0\%}  & 0.0\%          \\
\textbf{CMO VS CMO-Type}   & \textbf{18.3\%} & 2.8\%          & 8.5\%          & 5.6\%           & \textbf{12.5\%} & 9.7\%          & 11.1\%          & \textbf{12.5\%} & 9.7\%          & 2.7\%           & \textbf{4.0\%}  & 0.0\%          \\
\textbf{CMO VS CMO-MB}     & \textbf{14.1\%} & 8.5\%          & 8.5\%          & 4.2\%           & \textbf{9.2\%}  & 8.3\%          & 11.1\%          & \textbf{13.9\%} & 6.9\%          & 2.7\%           & \textbf{5.3\%}  & 0.0\%          \\ \hline
\end{tabular}
\end{table}

\section{Threats to Validity}
\label{sec:ttv}

In this section we list potential threats impacting the validity of our experiments. %However, it's possible that some measures might not have been effective.

\textbf{Construct Validity} The quality of the generated messages may have been affected by our prompt design. To ensure the quality of our prompts, we followed best practices \cite{awesomeprompt,ibm} and a trial-and-error process where multiple authors manually evaluated the quality of the optimized messages on a sample of commits. The superior performance of CMO also indicates the effectiveness of our approach.

Potential subjectivity in human evaluation was addressed by providing clear definitions of the metrics and examples to guide the process. Furthermore, multiple researchers and developers independently assessed the commit messages, ensuring a more objective evaluation.

%Potential subjectivity could be introduced in the human evaluation. In mitigate such subjectivity, we have provided detailed definitions of the evaluation metrics and demonstration examples to clarify the evaluation process. Also, the commit messages were evaluated independently by multiple researchers and developers. 

\textbf{Internal Validity} 
The dataset used in this study consists of only 381 commits, which may impact the performance of the \textit{LLM-based Quality Evaluator}, the effectiveness of CMO, and the evaluation results of the techniques. However, this dataset is the only one manually labeled with the evaluation metrics as explained in (Section \ref{sec:hm_metrics}) and was used to evaluate OMG. Therefore, using it ensures a fair comparison between OMG and CMO.

In addition, the identified human-considered contexts that are missed by OMG may not be the actual contexts that the authors of the commits considered when maintaining the projects. To mitigate this threat, two of the authors independently reviewed the code changes and their messages to draft the contexts possibly considered by developers. After that, they also closely examined the identified contexts to verify their necessity in composing the messages.

Due to financial constraints and the cost of using the OpenAI API, we could not exhaustively tune the of these hyper-parameters across all commits in the dataset. The selected of these hyper-parameters may not be fully optimal or generalizable. Nevertheless, we applied Grid Search for the of these hyper-parameters while evaluating the samples. Since CMO still outperforms OMG/CMC, it demonstrates the effectiveness of our tuning process.

%The size of the dataset adopted in this study is limited to only 381 commits, which may affect the performance of \textit{LLM-based Quality Evaluator}, the effectiveness of CMO, and the evaluation results of the studied techniques. However, the dataset is the only one that was manually labeled in the evaluation metrics (Section \ref{sec:hm_metrics}), on which OMG was evaluated. Using it would ensure a fair comparison between OMG and CMO. 

%In addition, due to financial restrictions and the cost of calling OpenAI API, we could not tune the hyper-parameters exhaustively on all commits in the dataset. The selected hyper-parameters might not be optimal and generalizable. However, CMO still outperforms OMG/CMC and significantly optimizes the human-written messages, which indicates the effectiveness of our tuning process.

\textbf{External Validity} Our findings may not be generalizable to all open source projects. Also, we selected GPT-4 because it has shown state-of-the-art performance in CMG \cite{li2024only}. Hence, the conclusions may not be generalized to other LLMs. 

\section{Conclusion and Future Work}
\label{sec:conclusion}
In this study, we identify software contexts considered by humans but missed by the state-of-the-art CMG technique, OMG. To address this, we propose \textit{Commit Message Optimization} (CMO), which leverages GPT-4 to optimize human-written messages by incorporating these missed contexts. Several automated quality evaluators serve as the objective function, aiming to maximize message quality. Our results demonstrate that CMO outperforms both OMG and CMC in terms of \textit{Rationality}, \textit{Comprehensiveness}, and \textit{Expressiveness}.

Future work could explore how enhancing the quality of the commit messages, while preserving human-considered information, impacts other SE tasks such as security patch identification \cite{zhou2021spi} and patch correctness prediction \cite{tian2022change}.

\bibliographystyle{ACM-Reference-Format}
\bibliography{fse}

\end{document}